\documentclass[12pt,preprint]{aastex}
\usepackage{psfig,epsfig,amsfonts,amsmath,amssymb,graphicx,morefloats,appendix,tensor}
\usepackage{color}
\usepackage{natbib}
\begin{document}

\slugcomment{Accepted to ApJ: October 26, 2018}

\title{ALMA Detection of Extended Millimeter Halos in the HD 32297 and HD 61005 Debris Disks}

\author{Meredith A. MacGregor\altaffilmark{1,2}, Alycia J. Weinberger\altaffilmark{1}, A. Meredith Hughes\altaffilmark{3}, D. J. Wilner\altaffilmark{4}, Thayne Currie\altaffilmark{5}, John H. Debes\altaffilmark{6}, Jessica K. Donaldson\altaffilmark{1}, Seth Redfield\altaffilmark{3}, Aki Roberge\altaffilmark{7}, Glenn Schneider\altaffilmark{8}}

\altaffiltext{1}{Department of Terrestrial Magnetism, Carnegie Institution for Science, 5241 Broad Branch Road NW, Washington, DC 20015, USA}
\altaffiltext{2}{NSF Astronomy and Astrophysics Postdoctoral Fellow}
\altaffiltext{3}{Department of Astronomy, Wesleyan University, Van Vleck Observatory, 96 Foss Hill Drive, Middletown, CT 06457, USA}
\altaffiltext{4}{Harvard-Smithsonian Center for Astrophysics, 60 Garden St., Cambridge, MA 02138, USA}
\altaffiltext{5}{National Astronomical Observatory of Japan, Subaru Telescope, National Institutes of Natural Sciences, Hilo, HI 96720, USA}
\altaffiltext{6}{Space Telescope Science Institute, 3700 San Martin Drive, Baltimore, MD, 21218, USA}
\altaffiltext{7}{Exoplanets and Stellar Astrophysics Lab, NASA Goddard Space Flight Center, Greenbelt, MD 20771, USA}
\altaffiltext{8}{Steward Observatory, The University of Arizona, 933 North Cherry Avenue, Tucson, AZ 85721, USA}

\begin{abstract}

We present ALMA 1.3 mm (230~GHz) observations of the HD~32297 and HD~61005 debris disks, two of the most iconic debris disks due to their dramatic swept-back wings seen in scattered light images.  These observations achieve sensitivities of 14 and 13~$\mu$Jy~beam$^{-1}$ for HD~32297 and HD~61005, respectively, and provide the highest resolution images of these two systems at millimeter wavelengths to date.  By adopting a MCMC modeling approach, we determine that both disks are best described by a two-component model consisting of a broad ($\Delta R/R> 0.4$) planetesimal belt with a rising surface density gradient, and a steeply falling outer halo aligned with the scattered light disk.  The inner and outer edges of the planetesimal belt are located at $78.5\pm8.1$~AU and $122\pm3$~AU for HD~32297, and $41.9\pm0.9$~AU and $67.0\pm0.5$~AU for HD~61005.  The halos extend to $440\pm32$~AU and $188\pm8$~AU, respectively.  We also detect $^{12}$CO J$=2-1$ gas emission from HD~32297 co-located with the dust continuum.  These new ALMA images provide observational evidence that larger, millimeter-sized grains may also populate the extended halos of these two disks previously thought to only be composed of small, micron-sized grains.  We discuss the implications of these results for potential shaping and sculpting mechanisms of asymmetric debris disks. 

\end{abstract}


\section{Introduction}
\label{sec:intro}

Debris disks are often seen as a sign of previous successful planet formation in a system \cite[see recent reviews including][]{hug18,wya18}.  Planetesimals, larger bodies leftover from the planet formation process, collide within these disks and grind down into smaller and smaller dust grains that are observable via scattered light or thermal emission.  It is expected that any interior planets should interact with this dusty material, sculpting it through gravitational perturbations.  In a few systems, we can link the geometry and location of a dust belt back to known planets.  For example, the presence of a giant planet was inferred from a warp in the $\beta$ Pictoris debris disk and later confirmed through direct imaging \citep{mou97,lag10}.  Given the known interactions, we could potentially use the observed structure of debris disks to place constraints on the orbits and masses of exoplanets that would otherwise be undetectable with current methods.  The HR~8799 system with four directly imaged planets between two dust belts offers a template case \cite[e.g.,][]{su09}; \cite{wil18} use the location of the outer disk's inner edge determined from millimeter observations to place a constraint on the mass of the outermost directly imaged planet.  In both $\beta$ Pictoris and HR~8799, the planet mass estimates derived from disk structure provide a separate check on limits derived from luminosity evolution, atmospheric modeling, and dynamical stability \citep{lag09,cur11}.  Unfortunately, this seemingly simple picture is complicated since structure in debris disks can be created through alternative, non-planetary mechanisms including stellar flybys, interactions with the interstellar medium \cite[ISM, e.g.,][]{art97}, recent massive collisions \cite[e.g.,][]{maz14}, or instabilities from dust and gas interactions \cite[e.g.,][]{lyra13}.

Scattered light imaging has revealed a population of edge-on debris disks that exhibit dramatic asymmetric structures including bowed disk midplanes (e.g., `wings' such as HD 32297 and HD 61005) and extensions (e.g., `needles' such as HD 15115).  To date, the dominant mechanism at work shaping these systems remains the subject of debate.  The leading theories involve interactions between disk material and either the ISM or an interior giant planet.  In the first scenario, ram pressure from interstellar gas can remove bound or unbound grains from the disk \citep{deb09,man09}.  On the other hand, an eccentric ($e>0.2$) giant planet orbiting within the disk can secularly perturb a narrow ring of parent bodies producing small dust grains that are thrown outwards via stellar radiation pressure \citep{esp16,lee16}.  A planet could also directly perturb grains over many orbits, forcing them to have high eccentricities and inclinations \citep{man09}.  

Since they both show wing-like features in scattered light observations, the HD~32297 (A5V, $<30$~Myr, $133\pm1$~pc) and HD~61005 (G8V, 40~Myr, $36.56^{+0.03}_{-0.06}$~pc, nicknamed `The Moth') debris disks are a useful pair to explore the mechanisms responsible for shaping asymmetric debris disks.  We note that the recent \emph{Gaia} Data Release 2 \citep{gaia16,gaia18} gives an updated distance for the HD~32297 system of 133~pc compared with the \emph{Hipparcos} result of 112~pc \citep{perry97}.  The difference between the \emph{Gaia} and \emph{Hipparcos} distances is smaller for HD~61005 with values of 36.6 and 34.5~pc, respectively.  Both of these disks were first resolved in scattered light at 1.1~$\mu$m with HST NICMOS \citep{sch05,hines07}, and have been subsequently imaged at millimeter wavelengths with low angular resolution \citep{man08,ric13,ste16}. We know that the small grains that dominate scattered light observations are more easily affected by non-gravitational forces such as stellar radiation and winds.  Larger, millimeter-sized grains are expected to be less affected by these interactions, and therefore more reliably trace the underlying planetesimal distribution in these disks.  Indeed, interactions with the ISM will preferentially strip small grains from the disk, but are unable to affect larger grains.  Thus, high resolution observations at millimeter wavelengths are especially critical to our understanding of the physical mechanisms shaping the structure of these asymmetric debris disks.

Here, we present new observations from the Atacama Large Millimeter/submillimeter Array (ALMA) with the highest angular resolution to date of both the HD~32297 and HD~61005 debris disks.  Our new ALMA images reveal that despite differences in spectral type, both systems are best described by a two-component structure with (1) a parent body belt, and (2) an outer halo coincident with the scattered light disk.  In Section~\ref{sec:obs}, we present the details of our observations.  In Section~\ref{sec:results}, we discuss the structure of the millimeter continuum emission for both disks (\ref{sec:cont}), the modeling approach we take to characterize the observed visibilities (\ref{sec:model}), and the detection of $^{12}$CO J$=2-1$ emission from the HD 32297 system (\ref{sec:co}).  In Section~\ref{sec:disc}, we compare our best-fit models to previous work (\ref{sec:comparison}).  We also explore the significance of our results in the context of debris disk shaping (\ref{sec:halo}) and stirring (\ref{sec:density}) mechanisms, and the origin of gas in debris disks (\ref{sec:gas}).  Finally, in Section~\ref{sec:conc}, we present our conclusions.

\section{Observations}
\label{sec:obs}

We observed both the HD~32297 and HD~61005 systems with ALMA in Band 6 (1.3~mm, 230~GHz, \#2015.1.00633.S, PI: Alycia Weinberger).  For HD~32297, ALMA executed two scheduling blocks (SBs) in Cycle 3 with 36 antennas in two configurations: (1) a 40-minute SB (22~minutes on-source) with baselines of $15-310$~m on 2016 January 1, and (2) a 68-minute SB (44~minutes on-source) with baselines of $15-704$~m on 2016 June 21.  For HD~61005, compact observations (baseline lengths of $12-335$~m) were taken previously in ALMA Cycle 1 (2012.1.00437.S, PI: David Rodriguez) on 2013 December 4 and 2014 March 20 with 27 and 31 antennas in the array, respectively.  Analysis of these lower resolution observations was presented in \cite{olof16}.  ALMA executed an additional 68-minute SB (44~minutes on-source) in Cycle 3 with 36 antennas in the array and baselines extending to 650~m on 2016 June 18.  Additional details (including precipitable water vapor and time on source) for all of these observations are listed in Table~\ref{tab:obs}.

The correlator for our new ALMA observations was set-up to maximize sensitivity to continuum dust emission.  For HD~32297, we also covered the $^{12}$CO J$=2-1$ line at 230.538~GHz with low spectral resolution (1.3~km~s$^{-1}$).  We used four basebands centered at 215.5, 217.5, 230.538, and 232.538~GHz, and two polarizations (XX and YY).  The baseband targeting the $^{12}$CO line had 3840 channels with a bandwith of 1.875~GHz.  The three basebands targeting only continuum emission included 128 channels with a bandwidth of 2~GHz.  The previous ALMA observations of HD~61005 included four basebands centered at 213, 215, 228, and 230.5~GHz.  The first three basebands targeted continuum emission with 128 channels and 2~GHz bandwidth, while the last baseband targeted the $^{12}$CO J$=2-1$ line using 3840 channels and a bandwidth of 1~GHz.  Since no CO emission was detected in the earlier ALMA observations, we did not include the $^{12}$CO line in our more extended observations of HD~61005, and included only four continuum basebands with a bandwidth of 2~GHz each centered at 224, 226, 240, and 242~GHz, in both the XX and YY polarizations.

All of the raw data sets were calibrated with the ALMA pipeline using the \texttt{CASA} package (version 4.7.2).  For HD~32297, observations of J0510+1800 were used for both bandpass and flux calibration.  Observations of J0502+0609 were used to account for time-dependent gain variations due to instrumental and atmospheric effects.  For HD~61005, observations of J0538-4405 were used for bandpass and flux calibration, while observations of J0747-3310 were used for gain calibration.  The systematic uncertainty for absolute flux calibration is estimated to be $<10\%$.  Visibility weights from ALMA are known to be inaccurate when reduced in versions of CASA prior to 4.4 \citep{CASA}.  This is not expected to be a problem with these data given that they were reduced using a more recent version of CASA.  However, to be safe, we used the \texttt{statwt} task in CASA to recalculate the data weights according to their true scatter such that the weights are consistent with $1/\sigma^2$, where $\sigma$ is the per-channel noise of the data, before proceeding with our analysis.  The calibrated visibilities were averaged in 30~second intervals to reduce the total size of the measurement sets, resulting in a total number of visibility points of 591,384 and 613,038 for HD~32297 and HD~61005, respectively.  To analyze the $^{12}$CO line emission from HD~32297, the continuum level was subtracted from the spectral window containing the line.  Both continuum and line images were generated using the multi-frequency synthesis \texttt{CLEAN} algorithm in the \texttt{CASA} package.

\section{Results and Analysis}
\label{sec:results}

\subsection{Continuum Emission}
\label{sec:cont}

\begin{figure}[t]
\begin{minipage}[h]{0.49\textwidth}
  \begin{center}
       \includegraphics[scale=0.75]{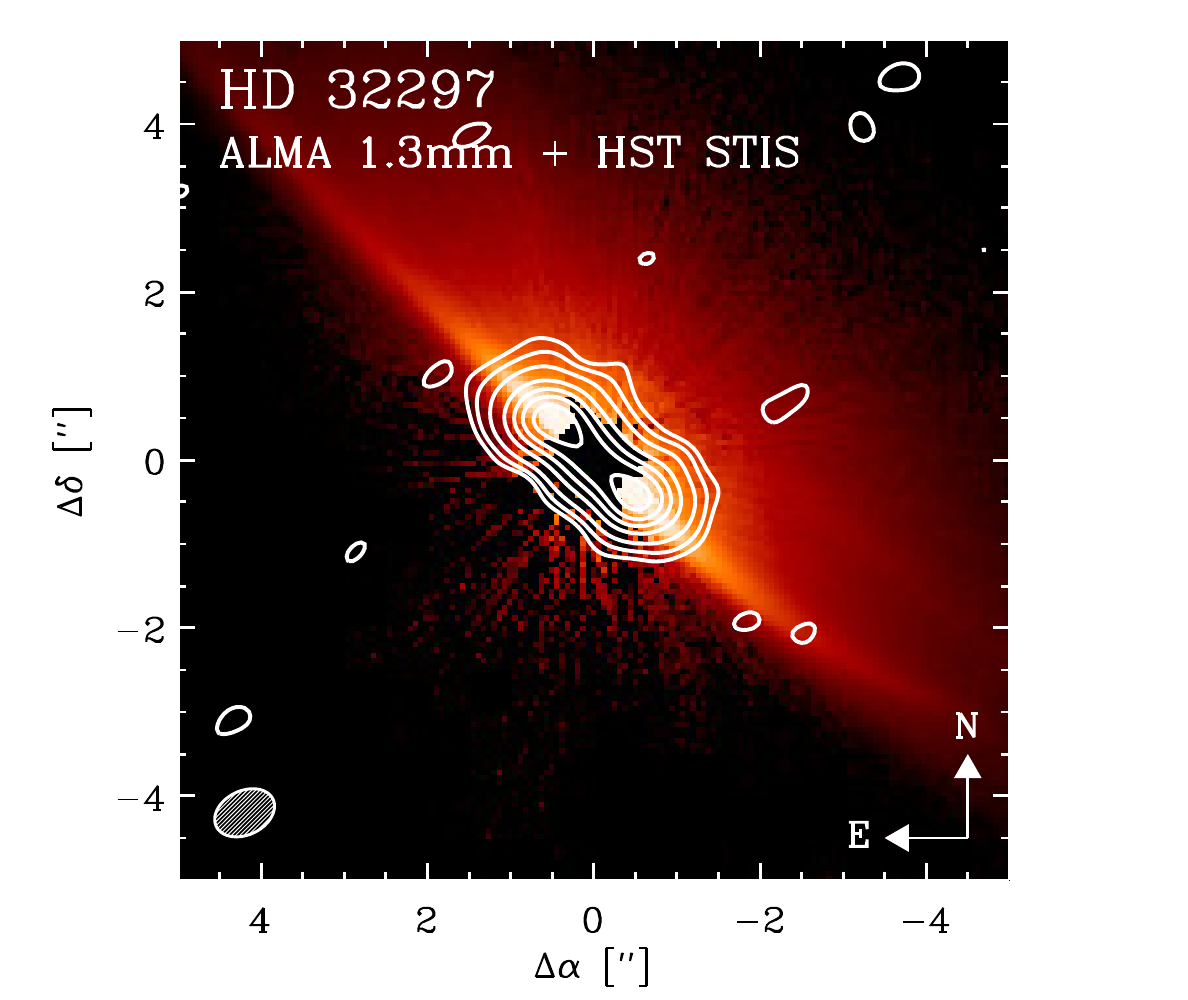}
  \end{center}
 \end{minipage}
\begin{minipage}[h]{0.49\textwidth}
  \begin{center}
       \includegraphics[scale=0.75]{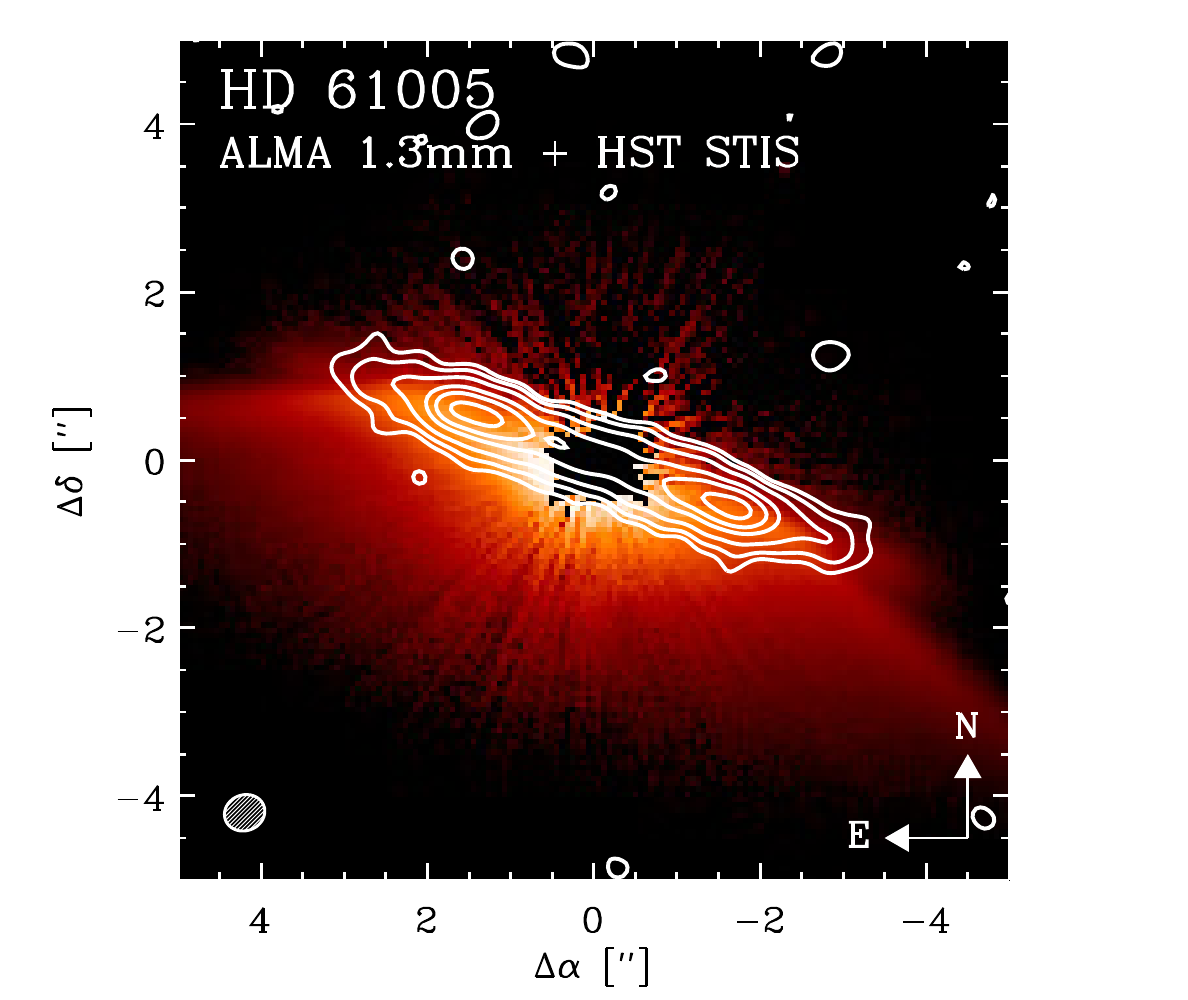}
  \end{center}
 \end{minipage}
\caption{\small New ALMA images of HD~32297 \emph{(left)} and HD~61005 \emph{(right)} match closely to previous scattered light observations, and show evidence for an additional halo component not seen before in lower resolution millimeter images.  The ALMA 1.3~mm continuum emission is overlaid as white contours on HST STIS images from \cite{sch14}.  The extended halo emission is seen in the $2-6\sigma$ contours beyond the disk ansae (the bright peaks on either side of the star that are characteristic of limb brightening from an edge-on, optically thin disk).  Contour levels in both panels are in steps of $[-3, 3, 6, 12, 24, 36, 48, 60]\times$ the rms noise of 14 and 13~$\mu$Jy for HD~32297 and HD~61005, respectively.  The dashed white ellipse in the lower left corner of both panels shows the beam size with robust$=0.5$ weighting of $0\farcs76\times0\farcs51$ and $0\farcs49\times0\farcs43$, again for HD~32297 and HD~61005 respectively.  
}
\label{fig:cont}
\end{figure}

Figure~\ref{fig:cont} shows our new ALMA 1.3~mm continuum images of HD~32297 (left) and HD~61005 (right) overlaid as contours on the previous \emph{Hubble Space Telescope} (HST) STIS coronagraphic images of optical scattered light from \cite{sch14}.  With robust $=0.5$ weighting, the synthesized beam size is $0\farcs76\times0\farcs51$ (100~AU $\times$ 68~AU at 133~pc, position angle $=-65\degr$) and the rms noise is 14~$\mu$Jy~beam$^{-1}$ for HD~32297.  For HD~61005, the synthesized beam size is $0\farcs49\times0\farcs43$ (18~AU $\times$ 16~AU at 36.6~pc, position angle $=-71\degr$) and the rms is comparable, 13~$\mu$Jy~beam$^{-1}$.

For both sources, the millimeter continuum emission aligns well with the orientation of the disk major axis seen in the scattered light images.  The HST images for these two disks are characterized by broad, wing-like structures extending to the northwest and southeast of the disk major axis for HD~32297 and HD~61005, respectively.  While we do not detect such large wing structures for either source in our ALMA images, both do show indications of some extended emission (seen in the $3-12\sigma$ contours) beyond the disk ansae (the bright peaks seen on either side of the star that are characteristic of limb brightening from an edge-on, optically thin disk), which we interpret as an outer~halo.  

The presence of significant millimeter emission at large stellocentric distances is atypical for debris disks, most of which appear sharply truncated at an outer radius.  For HD~61005, the extended halo emission matches well with the outer scattered light that is along the disk midplane.  The structure of the halo is somewhat less apparent by-eye for HD~32297, given its much larger distance (113~pc compared to 36.6~pc for HD~61005) and thus lower resolution. Despite this, these new ALMA images indicate that both the HD~32297 and HD~61005 debris disks are best characterized by a two component model with (1) a planetesimal belt characteristic of most millimeter observations of debris disks and (2) an additional outer halo extending $\sim2\arcsec-4\arcsec$ radially.

\subsection{Continuum Modeling Approach and Results}
\label{sec:model}

\begin{figure}[t]
\begin{minipage}[h]{0.25\textwidth}
  \begin{center}
       \includegraphics[scale=0.27]{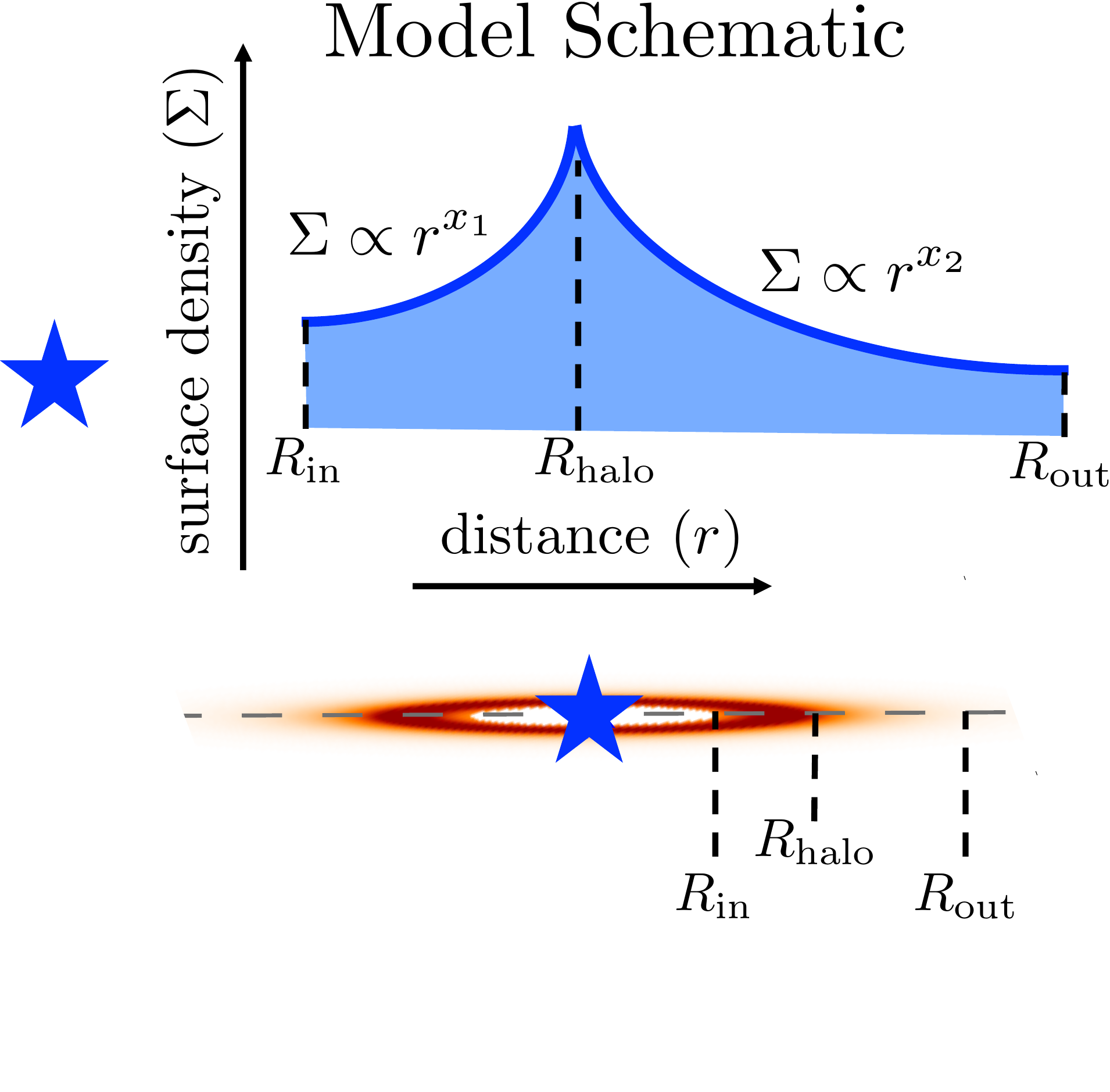}
  \end{center}
 \end{minipage}
\begin{minipage}[h]{0.75\textwidth}
  \begin{center}
       \includegraphics[scale=0.63]{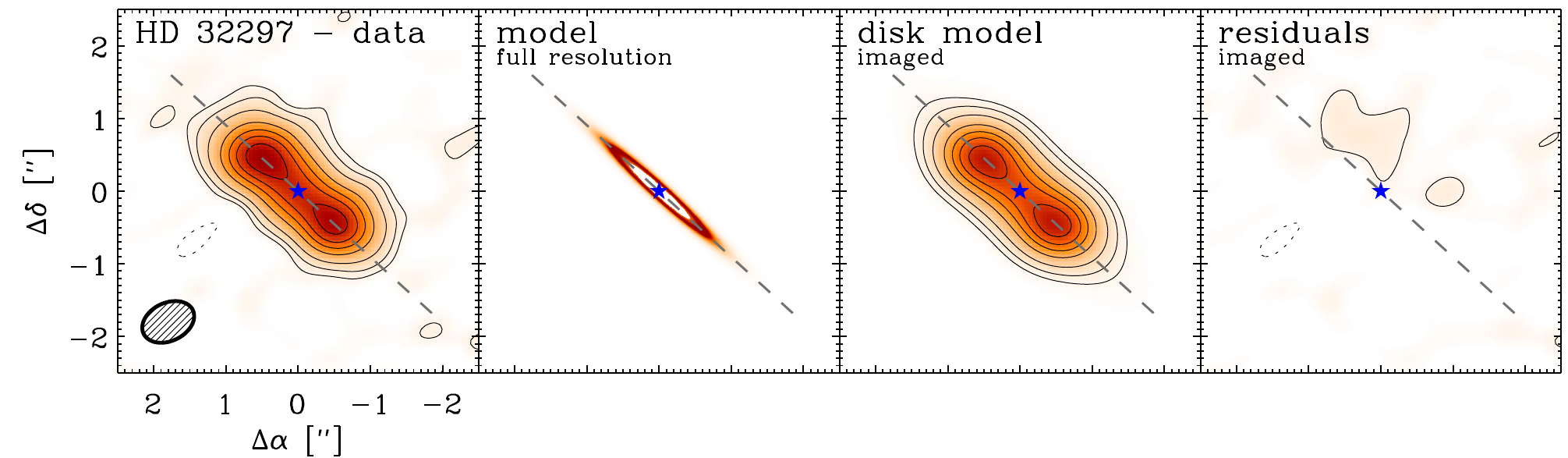}
       \includegraphics[scale=0.63]{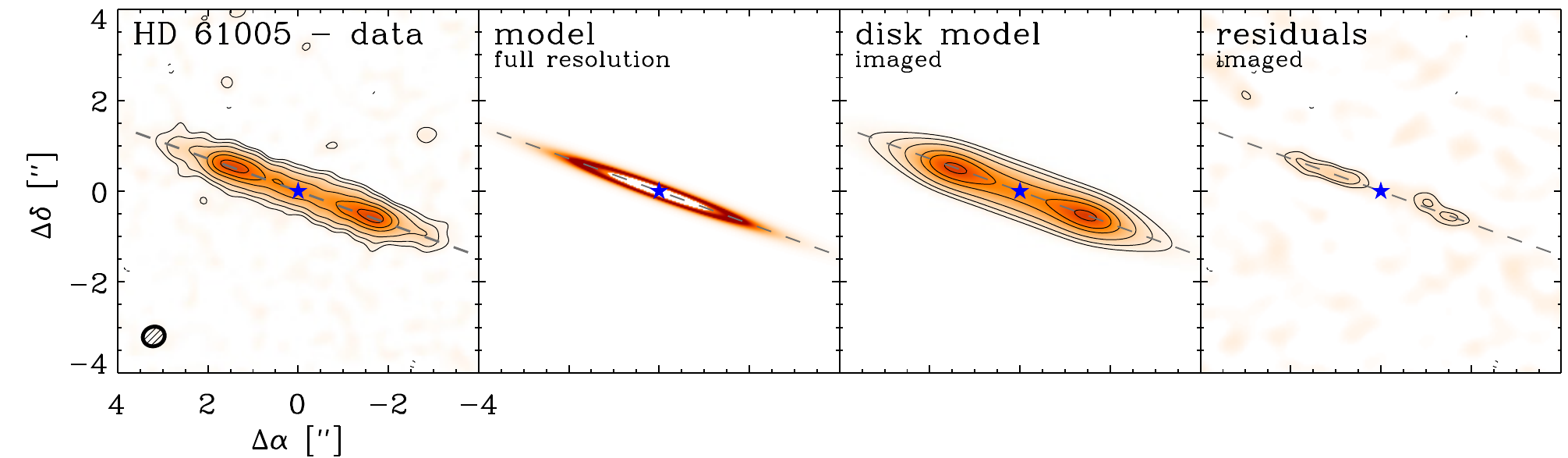}
  \end{center}
 \end{minipage}

\caption{\small A schematic of the two-component model surface density distribution is shown at left.  The best-fit model results (shown at right) for both HD 32297\emph{(top)}  and HD 61005 \emph{(bottom)} agree well with the data, although there are some $4-8\sigma$ residuals indicative of additional asymmetric structure in the disks not resolved by our current observations.  
For both top and bottom: \emph{(left)} The same ALMA 1.3~mm continuum map as is shown in Figure~\ref{fig:cont}.
\emph{(left, center)} The best-fit two-component model displayed at full resolution with a pixel scale of $0\farcs03$.
\emph{(right, center)} The same best-fit model imaged like the data with no noise.
\emph{(right)}  The resulting residuals again imaged like the data.  
Contours are in steps of $[-3, 3, 6, 12, 24, 36, 48, 60]\times$ the rms noise of 14 and 13~$\mu$Jy for HD~32297 and HD~61005, respectively.  The location of the central star is indicated by the star symbol, and the position angle of the disk major axis determined from model fitting is shown by the dashed gray line.  The synthesized beam, same size as in Figure~\ref{fig:cont}, is shown by the dashed ellipse in the leftmost panels.
}
\label{fig:dmr}
\end{figure}

We model the ALMA 1.3~mm continuum observations of both HD~32297 and HD~61005 with an optically thin, geometrically flat, axisymmetric belt model.  In order to draw comparisons between the two systems and because their scattered light and millimeter images show similar structures, we adopt the same model for both disks.  Model fits with only a single power law or Gaussian belt component \cite[as we have used previously for other debris disks including AU Mic and $\epsilon$ Eridani, see][]{mac13,mac15b} left significant residuals, most notably in the outer regions of the disk beyond the ansae.  We interpret these residuals as evidence for an outer halo component not accounted for by simple models with symmetric inner and outer edges.  Motivated by this apparent two-component structure of the ALMA continuum images (Figure~\ref{fig:cont}) of HD~32297 and HD~61005, we adopt a parametric model with a radial temperature profile, $T\propto r^{-0.5}$, and a surface density profile described by a broken power law with $\Sigma \propto r^{x_1}$ for $R_\text{in}<r<R_\text{halo}$ (within the planetesimal belt) and $\Sigma \propto r^{x_2}$ for $R_\text{halo}<r<R_\text{out}$ (within the halo).  A schematic of the model surface density distribution is shown in Figure~\ref{fig:dmr}.  

In our model fits, the inner radius of the planetesimal belt, $R_\text{in}$, outer radius of the planetesimal belt or inner radius of the `halo', $R_\text{halo}$, and outer radius of the halo, $R_\text{out}$, are all free parameters.  We fit for the flux of the planetesimal belt, $F_\text{belt}$, and halo component, $F_\text{halo}$, independently, with the total flux density normalized to $F_\text{tot}=F_\text{belt}+F_\text{halo}=\int I_\nu d\Omega$.  We assume that the surface density profile of both components is continuous such that $\Sigma_\text{belt}=\Sigma_\text{halo}$ at $R_\text{halo}$.  It is possible that the surface density profile could instead be discontinuous, which we do not consider here.  Adding a discontinuity introduces an additional model parameter, and the continuous model fits demonstrate that this is not necessary.  In addition, we fit for the belt geometry, namely the inclination, $i$, and position angle, $PA$.  Although we explored more complicated models that allowed for different surface density profiles, gaps between the belt and halo components, or multiple rings, they did not yield substantially different results.  In principle, other two-component parametric profiles, such as an asymmetric Gaussian, could be fitted to these data. We choose a broken power law as the simplest model justified by the sensitivity and resolution of our current dataset.

To obtain best-fit values and uncertainties for all parameters, we adopt our previously described approach \cite[e.g.,][]{mac13} and fit directly to the millimeter visibilities within a Markov Chain Monte Carlo (MCMC) framework making use of the \texttt{emcee} package \citep{for13}.  We compute synthetic visibilities for each model using the python package \texttt{vis\_sample}\footnote{\texttt{vis\_sample} is publicly available at \url{https://github.com/AstroChem/vis_sample} or in the Anaconda Cloud at
\url{https://anaconda.org/rloomis/vis_sample.}}, and then calculate a $\chi^2$ likelihood function, $\text{ln}\mathcal{L}=-\chi^2/2$ to compare with the data.  In order to fully explore the posterior probability distribution of the data conditioned on the model, we use $10^6$ iterations (100 walkers and 10,000 steps each).  To check for convergence, we compute the Gelman-Rubin statistic \citep{gel92} and ensure that $\hat{R}<1.1$ for all model parameters.  We assume uniform priors, with limits that ensure each model is physical:  $F_\text{belt}$, $F_\text{halo}$, $F_\text{tot}>0$ and $0<R_\text{in}<R_\text{halo}<R_\text{out}$.    

The best-fit parameter values and uncertainties for our model fits to the ALMA data of both HD~32297 and HD~61005 are presented in Table~\ref{tab:results}.  Both models provide reasonably good fits to the data with reduced $\chi^2$ values of 1.32 and 1.61, respectively.  Figure~\ref{fig:dmr} shows the ALMA 1.3~mm continuum image, the best-fit model at full resolution (pixel scale of $0\farcs03=$ 4~AU for HD~32297 and 1~AU for HD~61005), the best-fit model imaged like the data with no noise, and the resulting imaged residual visibilities for HD~32297 (top) and HD~61005 (bottom).  Although the best-fit models reproduce the bulk structure of both sources, some significant residuals are clearly seen.  For HD~32297, a $4\sigma$ residual peak is present on the northwest side of the disk major axis, on the same side as the swept-back wings seen in scattered light images.  We have chosen not to fit for vertical structure, given the moderate resolution of both datasets; there is no evidence that either disk is resolved in the vertical direction.  However, we note that given this choice, the residual emission from HD~32297 probably biases our model fits, pulling the best-fit value for the outer radius further away from the star (see Section~\ref{sec:compare_hd32297} for further discussion).  For HD~61005, residual emission ($\sim4\sigma$) is evident along the disk major axis between the disk ansae and the central star, with an $8\sigma$ peak located $\sim0\farcs5$ northeast of the stellar location (and present in all individual observations).  We cannot yet confirm whether this peak is also evident in scattered light images, since it lies at the edge of the HST STIS inner working angle ($\sim0\farcs4\sim14$~AU).  Both GPI and SPHERE have a small enough inner working angle to provide a direct comparison, and indeed, published SPHERE images show evidence for a brightness asymmetry \citep{olof16}.  Since residual emission is seen on both sides of the disk, one possibility is that this system contains an inner ring at $\sim18$~AU ($0.\!\!^{\prime\prime}5$ at a distance of 36.6~pc) that remains unresolved in our observations.  Multiple groups have modeled the spectral energy distribution (SED) of HD~61005, and concluded that there is strong evidence for an additional warm component in the system \cite[e.g.,][]{mor11,bal13,ric13,chen14,olof16}.  As of yet, we do not have direct imaging evidence for this claimed inner belt.  Additional observations with higher resolution at both millimeter and near-infrared wavelengths are needed to confirm the nature of this residual feature seen in our new ALMA image.

\begin{figure}[t]
\begin{minipage}[h]{0.49\textwidth}
  \begin{center}
       \includegraphics[scale=0.5]{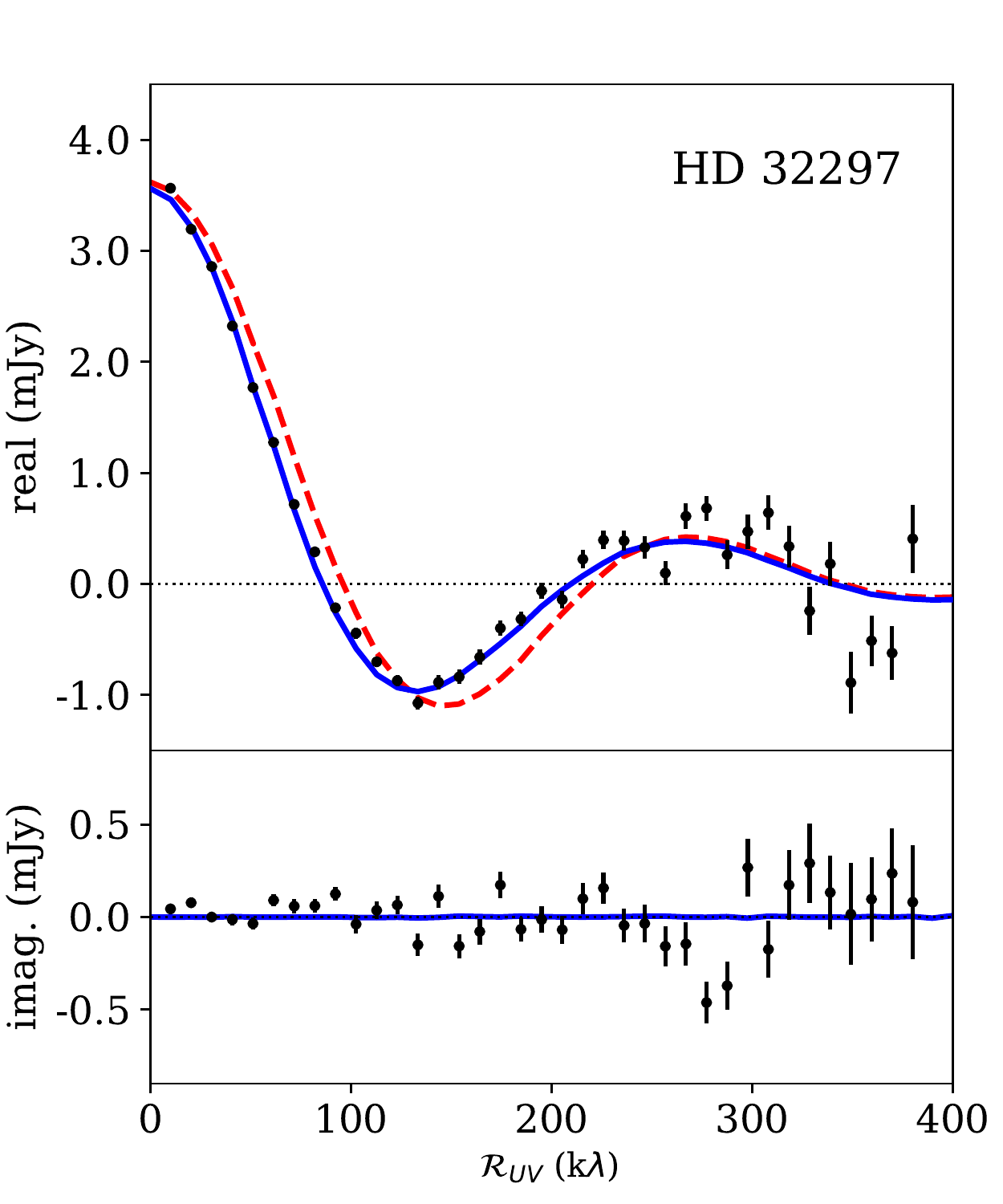}
  \end{center}
 \end{minipage}
\begin{minipage}[h]{0.49\textwidth}
  \begin{center}
       \includegraphics[scale=0.5]{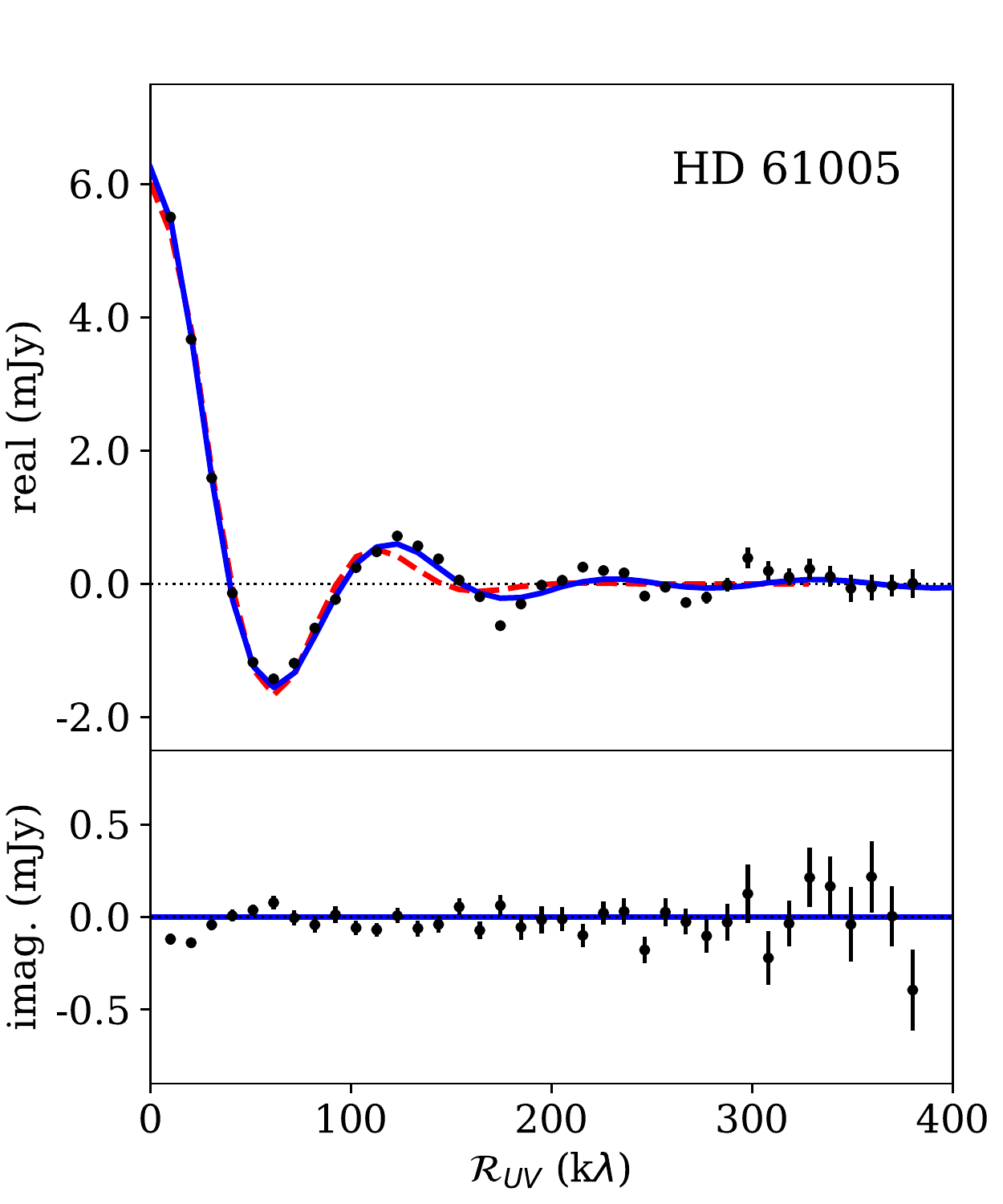}
  \end{center}
 \end{minipage}
 
\caption{\small The deprojected real and imaginary visibilities for HD~32297 \emph{(left)} and HD~61005 \emph{(right)}.  The 1.3~mm ALMA visibilities have been deprojected using the disk geometry determined by our model fits and averaged in bins of $10$~k$\lambda$.  In both plots, the solid blue line indicates the best-fit broken power law model, while the dashed red line shows a Gaussian fit to the data.  The Gaussian model without a halo component fails to fit the data well for either disk, especially in the first null and second peak ($50-250$~k$\lambda$) of the real visibility function.  For both sources, the imaginary visibilities show some low-amplitude structure indicative of asymmetric disk structure.
}
\label{fig:vis}
\end{figure}

We can also examine the quality of our model fits by looking at the deprojected visibilities.  Figure~\ref{fig:vis} shows the real (top) and imaginary (bottom) 1.3~mm visibilities deprojected using the disk geometry obtained from our model fits and averaged in concentric annular bins of $10$~k$\lambda$ in deprojected $(u,v)$ distance ($\mathcal{R}_{UV}$) for HD~32297 (left) and HD~61005 (right).  The best-fit broken power law models (solid blue line) agree very well with the data for $R_{UV}<150$~k$\lambda$, showing the characteristic peak and null of an annular belt of emission.  The data diverge from the models slightly at larger $(u,v)$ spacings, which could indicate the need for a more complicated, non-axisymmetric emission model.  For comparison, a Gaussian model fit to the data is shown as the dashed red line.  By eye, this simple model provides a poor fit to the real visibilities (especially in the first null and second peak of the visibility function from $50-250$~k$\lambda$), further motivating the need to include an additional component.  The imaginary visibilities provide the best probe of non-axisymmetric structure \cite[see discussion in][]{mac15b}.  In the case of an axisymmetric belt, the imaginary visibilities are expected to be uniformly zero.  For both HD~32297 and HD~61005, the imaginary visibilities instead show some low-amplitude structure that again hints at non-axisymmetric disk structure not resolved by our current observations.

\subsection{$^{12}$CO Emission}
\label{sec:co}

\begin{figure}[t]
\begin{minipage}[h]{0.49\textwidth}
  \begin{center}
       \includegraphics[scale=0.73]{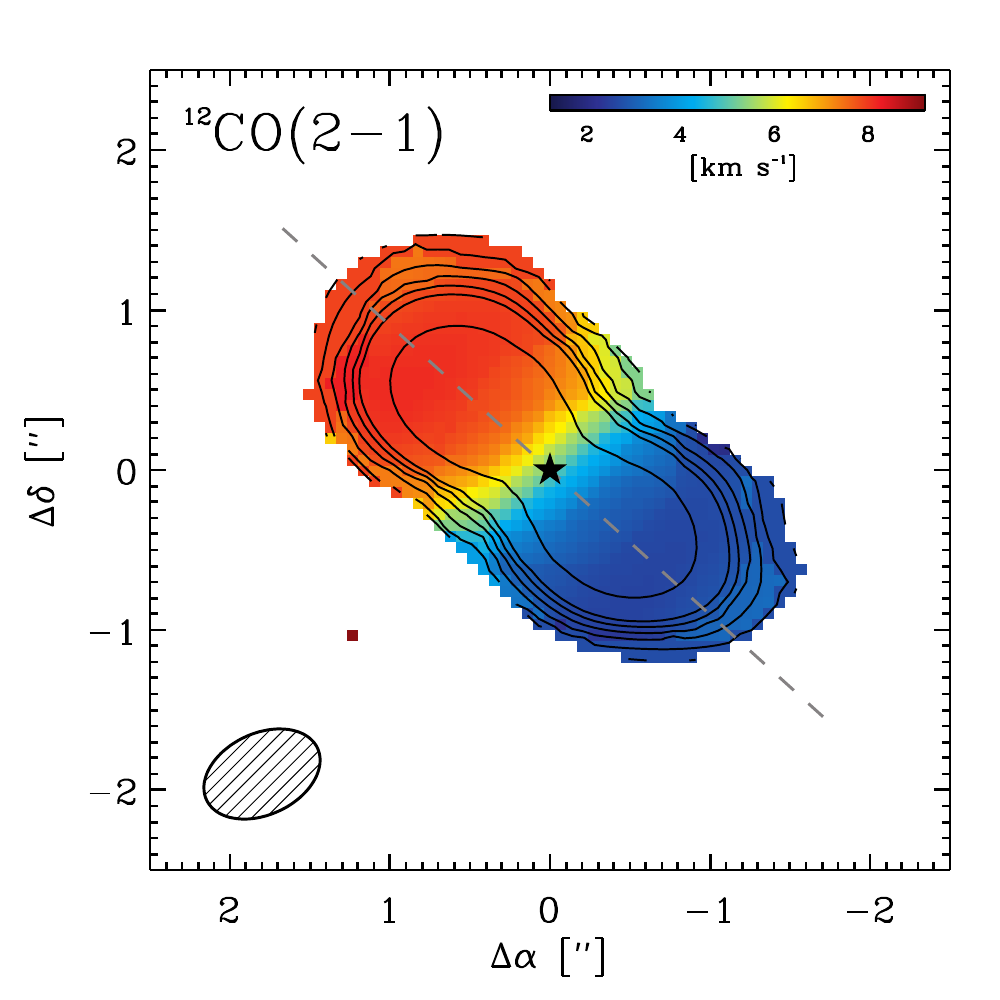}
  \end{center}
 \end{minipage}
\begin{minipage}[h]{0.49\textwidth}
  \begin{center}
       \includegraphics[scale=0.73]{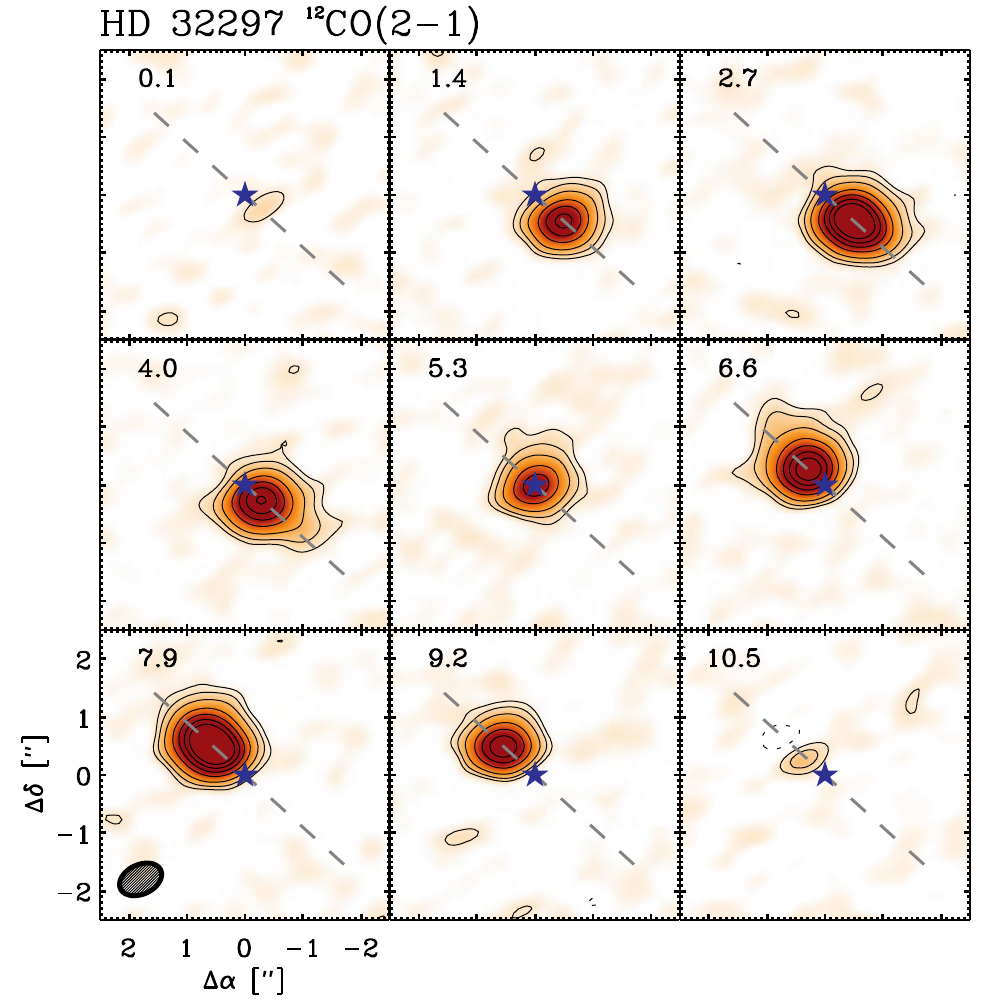}
  \end{center}
 \end{minipage}
\caption{\small We resolve the radial structure and rotation of the $^{12}$CO J$=2-1$ emission from the HD 32297 system with ALMA.  \emph{(left)} The $^{12}$CO J$=2-1$ moment maps for HD 32297.  Black contours show the 0th moment map (velocity-integrated intensity) in steps of $[3, 6, 12, 24, 36, 48, 60]\times1.0$~mJy~km~s$^{-1}$~beam$^{-1}$, the rms noise in the image, overlaid on the 1st moment (colors, intensity-weighted velocity).  A scale bar for the first moment is shown in the upper right corner for reference.  \emph{(right)}  The corresponding channel maps for $^{12}$CO.  Channels are 1.3~km~s$^{-1}$ wide with LSRK velocities labeled in the upper left corner of each panel.  Again, contours are in steps of $[3, 6, 12, 24, 36, 48, 60]\times1.0$~mJy~km~s$^{-1}$~beam$^{-1}$.  The leftmost panel in the bottom row shows the $0\farcs76\times0\farcs51$ synthesized beam size as a dashed ellipse.  In both figures, the star symbol indicates the location of the central star, and the dashed gray line shows the best-fit position angle of the disk major axis determined from modeling the continuum visibilities, PA$=47\fdg9$.
}
\label{fig:co}
\end{figure}

In addition to continuum dust emission, we also detect $^{12}$CO J$=2-1$ gas emission from the HD~32297 system.  Figure~\ref{fig:co} displays the moment (left) and channel maps (right), both of which show a clear velocity gradient that is indicative of rotation around the central star.  In the moment maps, the velocity-integrated intensity (0th moment) has been overlaid as contours on the intensity-weighted velocity (1st moment).  The central velocity in the barycentric frame is $20.6\pm0.3$~km~s$^{-1}$, consistent with the measured stellar radial velocity of $\sim23$~km~s$^{-1}$ \citep{tor06}.  Converting to the LSRK frame, which accounts for the peculiar motion of the Sun, yields a central velocity of $5.3\pm0.1$~km~s$^{-1}$.  The channel maps show only the central nine channels in which emission is clearly detected above $3\sigma$.  The rms noise with robust $=0.5$ weighting is 1.0~mJy~beam$^{-1}$.  The peak and integrated intensity are $0.28\pm0.003$~Jy~beam$^{-1}$ and $1.02\pm0.01$~Jy~km~s$^{-1}=7.91\pm0.08\times10^{-21}$~W~m$^{-2}$, respectively, determined over the region above $3\sigma$ in the 0th moment map.  No $^{12}$CO emission was detected from HD 61005 in the earlier Cycle 1 data, and \cite{olof16} placed an upper limit on the CO gas mass assuming Local Thermodynamic Equilibrium (LTE) of $7.7\times10^{-7}$~$M_\oplus$ (scaled to the new \emph{Gaia} distance, see Section~\ref{sec:intro}) for optically thin gas using these archival observations.  In contrast, \cite{kral17} predict a CO mass for HD~61005 in non-LTE of $1.8\times10^{-6}$~$M_\oplus$.

For HD~32297, we can calculate the total mass of CO probed by the $J=2-1$ line transition (noted by the subscript 21 in the following equation) for optically thin line emission in LTE:

\begin{equation}
    M_\text{CO} = \frac{4\pi}{h\nu_{21}} \frac{F_{21}md^2}{A_{21}x_{2}}
\end{equation}

\noindent Here, $F_{21}=7.91\times10^{-21}$~W~m$^2$ is the integrated flux density of the line, $m$ is the mass of the CO molecule, $d=133$~pc is the distance to the source, $h$ is Planck's constant, $\nu_{21}=230.538$~GHz is the rest frequency of the line, $A_{21}=6.91\times10^{-7}$~s$^{-1}$ is the Einstein A coefficient \citep{hol05}, and $x_2=\frac{N_2}{N_\text{tot}}$ is the fractional level population of the $J=2$ rotational level ($N_2$) relative to the total CO population ($N_\text{tot}$) evaluated at a characteristic temperature of 30~K.  This approach yields a total CO mass of $M_\text{CO} = 4.32\times10^{-4}~M_\oplus$.  If the CO were primordial and therefore accompanied by H$_2$, and assuming the canonical CO/H$_2$ ratio of $10^{-4}$ typical for the ISM, the total mass of molecular hydrogen in HD 32297 would be $\sim4.3$~$M_\oplus$.  

We have made two significant assumptions in our analysis of the $^{12}$CO emission from the HD~32297 debris disk, which may need to be reevaluated: (1) the gas is optically thin and (2) the disk is in LTE.  We can estimate the optical depth of the $J=2-1$ line from the 3-D data cube following \cite{mat17a} and again assuming LTE:

\begin{equation}
    \tau_\nu = \frac{d^2F_{21}}{A_{21}\Delta \nu \Delta A}\left(\frac{x_1}{x_2} B_{12}-B_{21}\right)
\end{equation}

\noindent where $d$ is the distance, $F_{21}$ is the same as above, $\Delta \nu$ is the line width, $\Delta A$ is the area of the emission on the sky along the line of sight (equivalent to the area of the $0\farcs76\times0\farcs51$ synthesized beam), $x_1$ and $x_2$ are the fractional populations of the lower and upper transition levels, and $A_{21}$, $B_{12}$, and $B_{21}$ are the Einstein coefficients.  To determine the fractional level populations, we assume a kinetic temperature of 30~K, appropriate for a belt at $\sim80$~AU.  From this calculation, we obtain a peak optical depth of $0.3$, reasonably consistent with optically thin emission.  However, previous observations of gas in debris disks  derive low excitation temperatures \cite[e.g.][]{kos13,fla16,hug17}, suggesting that the assumption of LTE might not be appropriate.  For non-LTE, a detailed radiative transfer model \cite[e.g., the Line
Modeling Engine or LIME,][]{bri10} would be required to obtain accurate gas masses.

\section{Discussion}
\label{sec:disc}

We have made new high resolution maps at millimeter wavelengths of the HD~32297 and HD~61005 debris disks with ALMA at 1.3~mm (230~GHz).  These two systems are notable for the dramatic swept-back wings evident in scattered light images.  Surprisingly, our new ALMA images hint at a more extended disk structure in millimeter emission that aligns with the scattered light wings.   Model fitting directly to the millimeter visibilities confirms that both disks are best described by a two-component structure with (1) a planetesimal belt, typical of debris disks observed at millimeter wavelengths, and (2) an extended outer halo, previously unseen in other millimeter images of these two systems with lower resolution.  Single-component models with either a power law or Gaussian surface brightness profile leave significant residuals in the outer regions of the disk.  Here, we compare our new observations to previous imaging studies, and discuss the implications for the origin of the halo component and stirring mechanisms.  Given our detection of $^{12}$CO J$=2-1$ emission from HD~32297, we also consider the properties of gas in debris disks.

\subsection{Comparison to Previous Observations}
\label{sec:comparison}

To place our new ALMA results into context, we compare our best-fit model structure and geometry for the HD~32297 and HD~61005 debris disks with previous imaging studies at shorter wavelengths.  Optical scattered light observations trace small, micron-sized grains rather than the larger, millimeter-sized grains we expect to dominate emission at ALMA wavelengths.  Comparing the disk geometry at these complementary wavelengths reveals the distribution of grain sizes within these disks.   In all of our model fitting, we have adopted the new \emph{Gaia} distances (discussed in Section~\ref{sec:intro}).  Thus, for the purposes of making a consistent comparison between our ALMA results and previous scattered light observations which assumed the \emph{Hipparcos} distance, we have scaled all previous determinations of radii for both disks to the new \emph{Gaia} distances.

\subsubsection{HD 32297}
\label{sec:compare_hd32297}

Our model fits place strong constraints on the radial location of the HD~32297 debris disk.  We note, however, that given the presence of residuals from the best-fit model (see Figure~\ref{fig:dmr}, it is likely that our quoted uncertainties are underestimates.  The inner edge of the planetesimal belt is located at $78.5\pm8.1$~AU with an outer edge at $122\pm3$~AU.  \cite{esp14} detect a break in the surface brightness profile of the disk at $\sim134$~AU in 1.6 and 2.2~$\mu$m Keck NIRC2 images.  Also using Keck NIRC2 coronagraphic imaging, \cite{cur12} see a similar break at $\sim130$~AU albeit with a more complicated surface brightness profile interior.  \cite{deb09} again find a break in HST images, but with an asymmetric radial location of 109 and 134~AU on the NE and SW sides of the disk, respectively. In the `birth ring' picture of debris disks, where small grains are produced through the collisional erosion of large planetesimals, we expect the outer edge of the planetesimal belt as traced by millimeter grains to line up with the break observed in the scattered light observations \cite[e.g.,][]{wil11}.  For HD~32297, there is good agreement between the outer edge of the planetesimal belt at 122~AU as seen by ALMA and the scattered light break observed by multiple telescopes at $\sim134$~AU.  Fits to the spectral energy distribution (SED) in the far-infrared confirm this radial structure, with a best-fit model that includes a cold dust ring centered at 134~AU \citep{don13}.  

Previous scattered light observations provide less information on the inner edge of the disk, mostly due to resolution challenges and large inner working angles.  \cite{moer07} note evidence for a dust depletion interior to $\sim85$~AU in mid-infrared observations at 11.7 and 18.3~$\mu$m.  \cite{fitz07} find that the 11.2~$\mu$m emission exhibits a bi-lobed structure characteristic of a disk with an inner edge at $\sim73$~AU.  Both of these inner edge locations are not well constrained with mid-infrared resolution of $\sim30-50$~AU at the distance of HD~32297, but the results are nonetheless similar to our best-fit model inner edge of $\sim79$~AU.  \cite{cur12} note a plateau in the surface brightness profile of the disk in Keck/NIRC2 imaging, which they interpret as evidence for two dust populations: (1) located between $\sim55-85$~AU and (2) truncated at $\sim133$~AU.  The radial locations of these two dust belts coincide with the inner edge of the planetesimal component and outer halo in our ALMA model fits.  HD~32297 was also imaged in the early days of CARMA at 1.3~mm \citep{man08}.  Combining these low resolution millimeter images with Spitzer MIPS photometry, \cite{man08} determine an inner edge location of $85^{+20}_{-10}$~AU, and also find evidence for two or three distinct grain populations within the disk.  Thus, the geometry of the main planetesimal belt is consistent across scattered light and millimeter wavelength imaging.  

Beyond the outer edge of the planetesimal belt at $122$~AU, the millimeter halo component extends to $440\pm32$~AU as determined from our ALMA observations.  As discussed in Section~\ref{sec:model}, these fits could be influenced by the residual emission on the NW side of the disk since we do not account for vertical structure in our models.  The uncertainty on this value is already quite large due to the low surface brightness in the outer regions of the disk in our model.  The degeneracy between fitting for an emission gradient and the outer radius of a power law disk model is well known \cite[e.g.,][]{mun96}.  The first HST images of the HD~32297 disk showed scattered light emission from a nearly edge-on disk out to $\sim480$~AU \citep{sch05} and nebulosity that extends out to $\sim2040$~AU, deviating from the midplane disk axis by $\sim15\degr$ on each side \citep{kal05}.  Using new HST/STIS broadband optical coronagraphy, \cite{sch14} find a similar extent of $\sim1890$~AU for the disk.  It is clear that the scattered light emission from small grains has a much larger radial extent and deviation from the midplane axis than the millimeter emission probed by ALMA. 

The millimeter disk geometry is in relatively good agreement with previous measurements of the midplane disk geometry from scattered light observations.  Our best-fit inclination and position angle are $83\fdg6 (+4\fdg6, -0\fdg4)$ and $47\fdg9\pm0\fdg2$, respectively.  \cite{sch14} find a position angle of $47\fdg5\pm1\fdg0$ from HST images, nearly identical to our result.  Previous determinations of the disk inclination range from $\sim80\degr$ \citep{kal05} to $\sim88\degr$ \citep{boc12}, overlapping with our result.  However, more recent scattered light images have preferred inclinations closer to $\sim88\degr$ \cite[e.g.,][]{boc12,cur12}, on the upper end of this range.  We attribute this discrepancy to the $4\sigma$ residuals on the northwest side of the disk major axis seen in Figure~\ref{fig:dmr}.  Our model fits favor lower inclinations in order to try and match this excess emission, and the asymmetric uncertainty on our reported best-fit inclination accounts for this discrepancy.  From a simple $\chi^2$ analysis in the image plane, we estimate an additional $5\%$ uncertainty towards higher inclinations, consistent with scattered light determinations.  If we subtract a model with a higher inclination (i.e., $88\degr$), we are left with larger residuals on the northwest side of the disk in excess of $8\sigma$, while no additional residuals are produced on the southeast side of the disk.  This finding indicates that there is significant millimeter emission towards the northwest of the disk major axis coincident with the extended scattered light wings that our current symmetric model does not account for (first discussed in Section~\ref{sec:model}). 

\pagebreak

\subsubsection{HD 61005}
\label{sec:compare_hd61005}

The best-fit model to the ALMA millimeter emission for HD~61005 has a planetesimal belt with an inner edge of $41.9\pm0.9$~AU and an outer edge of $67.0\pm0.5$~AU.  Again, we note that the uncertainties quoted here are likely underestimates given the significant residuals from the best-fit model (see Figure~\ref{fig:dmr}).  Gemini Planet Imager (GPI) total intensity observations at 1.6~$\mu$m probed down to projected separations of $<10$~AU, revealing a sharp disk inner edge at $\sim50$~AU \citep{esp16}, in agreement with the inner edge determined by ALMA of $\sim42$~AU.  Using complementary Keck NIRC2 angular differential imaging of the disk at $1.2-2.3$~$\mu$m, \cite{esp16} note asymmetric breaks in the disk surface brightness profile beyond 55 and 65~AU on the NE and SW sides of the disk, respectively.  As with HD~32297, the coincidence of the break observed in the scattered light surface brightness profile and the outer edge of the planetesimal belt determined by ALMA supports the conclusion that the observed small grains are produced through a collisional cascade occurring within a birth ring of planetesimals.

In contrast to HD~32297, the HD~61005 system has been observed multiple times at millimeter wavelengths.  \cite{ric13} first resolved the disk with the Submillimeter Array (SMA), and determined an inner edge for the disk of $71\pm 2$~AU.  A new analysis of the same SMA data by \cite{ste16} yielded a consistent result of $75.4^{+3.0}_{-4.7}$~AU.  \cite{olof16} fit models to the low resolution ALMA data included in this work (2012.1.00437.S, PI: David Rodriguez), and obtain a similar value for the disk reference radius of $70.6^{+6.1}_{-8.7}$~AU. Both of the SMA determinations place the inner disk edge farther out than our best-fit result, likely due to differences in model formulation and to the low angular resolution of these observations.  The reference radius determined by \cite{olof16} indicates the location of a power law break in the disk model, and is thus most comparable to our planetesimal outer belt edge of $\sim67$~AU. None of these previous millimeter images show any indication of the extended wing structures seen in scattered light observations.

The halo component, as newly detected by the higher resolution ALMA observations presented here, extends from the outer edge of the planetesimal belt at $67$~AU to $188\pm8$~AU.  \cite{esp16} detect swept-back wings with a deflection angle of $\sim22\degr$ in the KECK NIRC2 images, but they are outside of the GPI field of view.  The wings extend from $\sim62$ to 135~AU on the NE side of the disk, and from $\sim67$ to 143~AU on the SW side.  \cite{bue10} detect emission out to $\sim150$~AU with VLT/NACO, and HST images show extended emission at distances of $\gtrsim200$~AU from the star \citep{hines07,sch14}.  While the ALMA emission does not exhibit the significant deflection seen in scattered light observations, the extent of the halo component is similar at both wavelengths.  The geometry of the planetesimal belt is also in good agreement.  We determine a best-fit inclination of $85\fdg6\pm0\fdg1$ and position angle of $70\fdg3\pm0\fdg1$. \cite{sch14} determine a nearly identical inclination and position angle for the disk major axis of $85\fdg1$ and $70\fdg3$, respectively.

\subsection{On the Origin of the Millimeter Halo Component}
\label{sec:halo}

Since the wing-like structures of the HD~32297 and HD~61005 debris disks have previously only been seen in scattered light images, discussions of their origin have centered on mechanisms that affect predominantly small, micron-sized grains.  The millimeter-sized grains that dominate emission at ALMA wavelengths are expected to closely trace the location of the parent bodies in the disk.  Thus, we have typically assumed that these large grains should not populate extended halos.    Few previous observations have challenged this assumption, with the exception of \cite{mar16} who detected low-level extended emission from HD~181327, a face-on debris disk, which they interpreted as either a halo or unresolved second ring.  If the population of larger grains in HD~32297 and HD~61005 can be more extended, as our new ALMA observations suggest, the mechanisms invoked to produce the morphology of these disks must be re-evaluated.  In this section, we present several potential mechanisms and discuss their applicability to millimeter grains.  

It is possible that the millimeter emission detected in the halo components of HD~32297 and HD~61005 could result from an abundance of micron-sized grains instead of millimeter-sized  grains.  To assess the likelihood of this scenario, we can estimate the mass of small grains needed in the halo to account for the observed millimeter flux.  For optically thin dust emission, the dust mass is $M_\text{dust}=F_\nu D^2/(\kappa_\nu B_\nu(T_\text{dust}))$, where $F_\nu$ is the flux density, $D$ is the distance (133~pc for HD~32297, 36.6~pc for HD~61005), $\kappa_\nu$ is the dust opacity, and $B_\nu$ is the Planck function at the dust temperature, $T_\text{dust}$.  The main source of uncertainty in this calculation is the dust opacity.  For millimeter-sized grains, a commonly used value in the literature is $\kappa_\nu = 2.3$~cm$^2$~g$^{-1}$ at 1.3 mm \citep{beck90}.  In comparison, \cite{pol94} determine a dust opacity of $7.9\times10^{-3}$~cm$^2$~g$^{-1}$ for their 300~$\mu$m radius composite (50\%) grains and $4.4\times10^{-3}$~cm$^2$~g$^{-1}$ for 3~$\mu$m grains with the same composition, several orders of magnitude smaller.  In order to be able to compare to literature values for other debris disk systems, we adopt the typical value of $2.3$~cm$^2$~g$^{-1}$ in our calculations for millimeter-sized grains in the planetesimal belt, and $1.3$~cm$^2$~g$^{-1}$ for micron-sized grains in the halo, determined using the same ratio between small and large grains as the \cite{pol94} values.  These values are similar to those shown in \cite{and15} for grains that are a mixture of amorphous silicates, carbon, and vacuum with a filling factor of 0.5.  We note, however, that if we instead used the \cite{pol94} values, our calculated dust masses would be significantly higher.  

To obtain an estimate for the dust temperature, we assume radiative equilibrium and use the radial location from our best-fit models.  Since small grains are likely hotter than the balckbody equilibrium temperature, we adopt a simplified emission efficiency, $Q_\lambda = 1-e^-(\frac{\lambda}{2\pi a})^{-\beta}$, where $a$ is the grain size and $\beta$ is the power law index of grain emission efficiency at long wavelengths which we assume to be 0.7, comparable with measurements from other disks around main sequence stars \citep{wil04}.  For HD~32297, the best fit inner edge of the planetesimal belt is $78.5\pm8.1$~AU and the start of the halo component is $122\pm3$~AU, which gives dust temperatures of $\sim47$~K and $\sim38$~K for each component, respectively.  The total flux density of the HD~32297 planetesimal belt is $3.04\pm0.21$~mJy, which yields a total mass of millimeter-sized grains of $0.57\pm0.07$~$M_\oplus$.  The halo component has a total flux density of $0.63\pm0.26$~mJy and a total mass of micron-sized grains of $0.29\pm0.06$~$M_\oplus$.  For HD~61005, the planetesimal belt inner edge is at $41.9\pm0.9$~AU and the halo component starts at $67.0\pm0.5$~AU, implying dust temperatures of $\sim37$~K and $\sim29$~K, respectively.  The total flux density of the planetesimal belt is $4.82\pm0.29$~mJy yielding a total mass of millimeter-sized grains of $0.088\pm0.010$~$M_\oplus$, and the total flux density of the halo is $1.54\pm0.48$~mJy for a mass of micron-sized grains of $0.070\pm0.012$~$M_\oplus$.  We can then calculate the ratio of halo dust mass to planetesimal belt dust mass to be $0.51$ and $0.79$ for HD~32297 and HD~61005, respectively, a significant fraction for both systems.  

Recent work has shown that such high opacities may be unrealistic for micron-sized astro-silicate grains  in the Rayleigh regime \citep{kat14}.  If we assume a much lower opacity of $10^{-3}$~cm$^2$~g$^{-1}$, the implied mass fraction of the halo compared to the planetesimal belt is instead $800$ and $500$ for HD~32297 and HD~61005, respectively.  Mass ratios of $0.5-800$ (the full span of the opacity values we have considered here) are highly unphysical, and inconsistent with collisional evolution models.  Thus, it is unlikely that a large enough mass of such grains will build up to explain the observed millimeter fluxes, disfavoring any proposed mechanisms that operate preferentially on small grains.  A population of larger bodies with longer collisional lifetimes and higher opacities is needed in the halo of both disks.

One mechanism that operates predominantly on small grains and that has been discussed extensively in the literature \cite[e.g.,][]{art97,deb09,man09,pas17} is interactions with the local interstellar medium (ISM), in which disk grains interact with gas in an ISM cloud, and are subsequently blown out due to gas drag or ram pressure stripping.  Millimeter grains are not expected to feel a significant effect from interactions with the ISM.  If the grains are highly porous as indicated for some debris disks \citep[e.g., AU Mic, ][]{gra07}, it is possible that ISM interactions could still play a role, but further investigation would be required to better evaluate this scenario.  It has also been suggested that the observed halos could result from a density enhancement of small grains due to a recent massive collision, as has been invoked previously to explain asymmetric structure in the HD~15115 \citep{maz14} and $\beta$ Pictoris \citep{dent14} debris disks. Such an event would likely produce an asymmetric distribution of small grains and is expected to be transient \cite[lasting for $\sim1000$ orbits or a few Myr at a radius of $\sim50$~AU,][]{jack14}, neither of which is characteristic of the HD~32297 and HD~61005 systems.  

Interactions with massive planets are often invoked to explain observed disk structures, including asymmetries, warps, and offsets, and are able to sculpt both small and large grains.  To date, there are no detected planets in either the HD~32297 or HD~61005 systems.  However, scattered light imaging of both disks shows brightness asymmetries that could be indicative of eccentric structure.  \cite{cur12} report that the southwest side of the HD~32297 debris disk is $50-100\%$ brighter than the northeast side in Keck/NIRC2 coronagraphic images, which is mirrored in the early CARMA image at 1.3~mm \citep{man08}.  HST NICMOS and ACS observations of HD~61005 also show a brightness asymmetry, with the northeast side of the disk nearly twice as bright as the southwest \citep{hines07}. VLT NaCo \citep{bue10} and SPHERE \citep{olof16} observations of HD~61005 show an offset of the disk center to the west of the stellar position along the disk major axis giving an eccentricity for the disk of $e=0.045\pm0.015$ and $e\sim0.1$, respectively.  The significant residuals that we see in our new ALMA observations (see Section~\ref{sec:model} for discussion) further support the claim that the HD~61005 debris disk may be asymmetric.  These ALMA observations do not detect the central star or any asymmetry between the two disk ansae for either HD~32297 and HD~61005, which would be required to place strong constraints on any disk eccentricity.  Without detecting the star, we cannot determine whether there is an offset between the stellar position and the disk centroid.  Without detecting a brightness asymmetry, we cannot make any claims of pericenter or apocenter glow.  In an eccentric disk, the pericenter side of the disk is closer to the star and as a result glows more brightly at shorter wavelengths producing pericenter glow.  At longer (millimeter and radio) wavelengths, the apocenter side of the disk appears brighter due to a surface density enhancement \citep{pan16,mac17}. We did not attempt to model such a brightness asymmetry here, because it is not apparent in the residuals resulting from our simpler model.  In principle, the argument of pericenter is unconstrained by these observations; the pericenter of the disk could be located along the observed disk major axis or be projected along the line of sight.

To date, multiple models of secular perturbation by a planet have been proposed to explain the unusual morphology of the HD~32297 and HD~61005 debris disks.  \cite{esp16} consider a planet interior to the HD~61005 disk with an eccentricity of at least $0.2$, are able to produce models that exhibit the same features as scattered light observations.  A range of possible semi-major axes and masses are allowed, from an Earth-mass planet farther from the star at 35~AU to a Jupiter-mass planet close-in at 5~AU.  \cite{lee16} construct a more generalized debris disk model in which a possibly eccentric planet secularly perturbs a narrow ring of parent bodies producing small dust grains that are thrown outwards via stellar radiation pressure.  The resulting scattered light disk morphology depends on several factors, including the planet eccentricity, disk inclination, and whether or not the dust grains are on apsidally aligned orbits.  By varying these parameters, they are able to qualitatively reproduce the observed morphologies of both HD~32297 and HD~61005.  To produce the `moth'-like shape of HD~61005, a highly eccentric planet ($e\sim0.7$) is needed along with grains on orbits that are apsidally aligned with the parent ring and with their apoastra directed towards the observer.  The double wings of HD~32297 can be produced by a similar model, but with dust grains only launched at the parent body periastra.  Both of these models treat only the small grains that dominate scattered light images, but their results could be extended in order to compare to millimeter images.  \cite{man09} discuss a slightly different process by which a planet could perturb the orbits of disk grains over time, slowly forcing them to have high orbital eccentricities and inclinations.  It is also possible that an exterior planet could be responsible for sculpting these systems.  In the models of \cite{the14}, a planet exterior to an initially narrow ring produces a population of large grains outside of the planet's orbit.  Although these grains are eventually removed from the system, their lifetime is long enough to contribute to the optical depth of the disk at large radii.

Our new ALMA observations indicate that millimeter-sized grains are present in the halos of the HD~32297 and HD~61005 debris disks.  Since many of the previously proposed mechanisms operate predominantly on small grains, it is likely that interactions with planetary perturbers play a role in generating the wing structures seen at both optical and millimeter wavelengths.  In the case of HD~32297, the high mass of gas within the disk may also contribute to the dynamics of the system.  Ultimately, observations with higher angular resolution will reveal any asymmetric structure or disk centroid offsets that might be indicative of planetary sculpting in these disks.  If an eccentric interior planet really is responsible for shaping these systems, we would expect to see additional substructure in the inner regions of the disk as a result of its gravitational interaction with disk material.

\subsection{Constraints on Stirring Mechanisms from the Surface Density Gradient}
\label{sec:density}

Our model fits constrain the disk radial surface density profile, $\Sigma \propto r^{x}$, with power law index $x_1$ within the planetesimal belt and $x_2$ for the halo.  Both HD~32297 and HD~61005 exhibit a rising surface density profile for the planetesimal belt with $x_1=2.23\pm0.61$ and $2.87\pm0.19$, respectively.  The surface density falls off steeply in the halo as $x_2=-6.20\pm0.30$ for HD~32297 and $-5.52\pm0.14$ for HD~61005.  Resolved millimeter observations have placed constraints on the surface density gradients of a handful of other debris disks.  Notably, a growing population exhibit rising surface density gradients \cite[e.g., AU Mic and $\epsilon$ Eridani, see][]{mac15b,mac13}.  Intriguingly, many of these disks appear radially broad like HD~32297 and HD~61005, which have fractional widths ($\Delta R/R$) of $0.43\pm0.04$ and $0.46\pm0.01$, respectively. For AU Mic, $\Delta R/R=1.28^{+0.13}_{-0.01}$, and for $\epsilon$ Eridani, $\Delta R/R=0.31^{+0.09}_{-0.13}$.  In comparison, the fractional width of the confined Fomalhaut debris disk is only $0.10\pm0.01$ \citep{mac17}.  Some of these systems may in fact exhibit more complicated radial structures.  ALMA observations of the HD~107146 debris disk also indicate a rising surface density gradient, but with multiple concentric rings \citep{ric15}.    

Measuring the surface density gradient of debris disks could provide a key observational constraint on the mechanisms operating in these systems to incite collisions.  One possibility is that debris disks are `planet-stirred' through the gravitational influence of giant planets interior to the disk \citep{mus09}.  \cite{kenyon02,kenyon08} discuss an alternative `self-stirred' model in which Pluto-sized bodies (radii $\sim1000$~km) are produced on the inner edge of the disk and trigger a collisional cascade.  The growth time in a disk is proportional to $P/\Sigma$, where $P\propto r^{3/2}$ and $\Sigma$ is the initial surface density.  If the initial surface density decreases with radius, as is expected in a primordial disk, the growth time is shorter on the inner edge of the disk. For a disk with total dust mass comparable to HD~32297 and HD~61005, these models predict timescales of tens of Myr to form large bodies on the inner edge of a disk at tens of AU.  On the outer edge of the disk ($>100$~AU), the timescale is much longer, $\gtrsim1$~Gyr.  \cite{ken10} predict the resulting surface density profile in such a `self-stirred' disk, finding $\Sigma\propto r^{+7/3}$.  The ages of HD~32297 and HD~61005 are estimated to be $\sim30-40$~Myr.  At face value, this could allow enough time for larger bodies to form near the inner edge of the HD~61005 disk at $\sim42$~AU, but is not entirely consistent for HD~32297 with an inner edge nearly twice as far from the star at $\sim79$AU.  

There a few ways to solve this slight discrepancy in timescales.  The age of the HD~32297 system is not well-constrained.  Although stellar kinematics and the high disk luminosity suggest a young age ($\sim10-20$~Myr), evolutionary tracks allow from ages up to $\sim500$~Myr \citep{rod14}, which may provide more time to form large bodies on the inner edge of the disk.  In addition, recent work by \cite{kri18} considers the possibility that smaller (a few km to $\sim200$~km in size) planetesimals could effectively stir debris disks.  Such bodies form much more rapidly in protoplanetary disks through pebble concentration than larger Pluto-sized bodies.  If large bodies are in fact not needed to incite collisions in debris disks, both the HD~32297 and HD~61005 debris disks could be self-stirred.

Of course, it is possible that an entirely different stirring mechanism may be at work in the gas-rich HD~32297 disk.  There is some similarity between the surface density profile of this system and 49 Ceti, a comparable $\sim$40~Myr-old gas-rich debris disk.  Resolved ALMA observations of this disk are well-fit with a surface density profile that rises with a power law index of $1.44$ in the inner part of the disk between 7 and 68~AU, and then falls off with an index of $-2.5$ out to 215~AU \citep{hug17}.  High resolution (sub)millimeter observations of a larger population of gas-rich debris disks are needed to better understand the stirring mechanisms in these systems.

\subsection{Gas in the HD~32297 Debris Disk}
\label{sec:gas}

\begin{figure}[t]
\begin{minipage}[h]{0.45\textwidth}
  \begin{center}
       \includegraphics[scale=0.73]{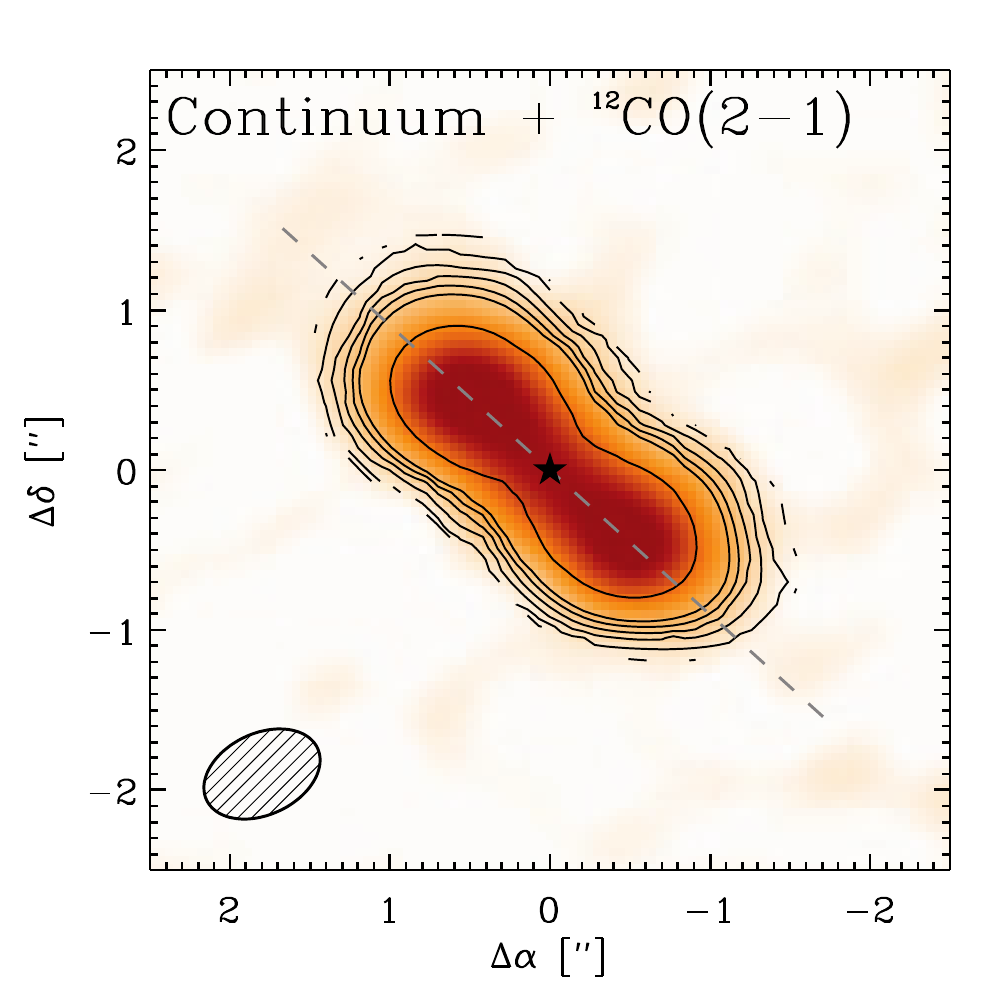}
  \end{center}
 \end{minipage}
\begin{minipage}[h]{0.55\textwidth}
  \begin{center}
       \includegraphics[scale=0.47]{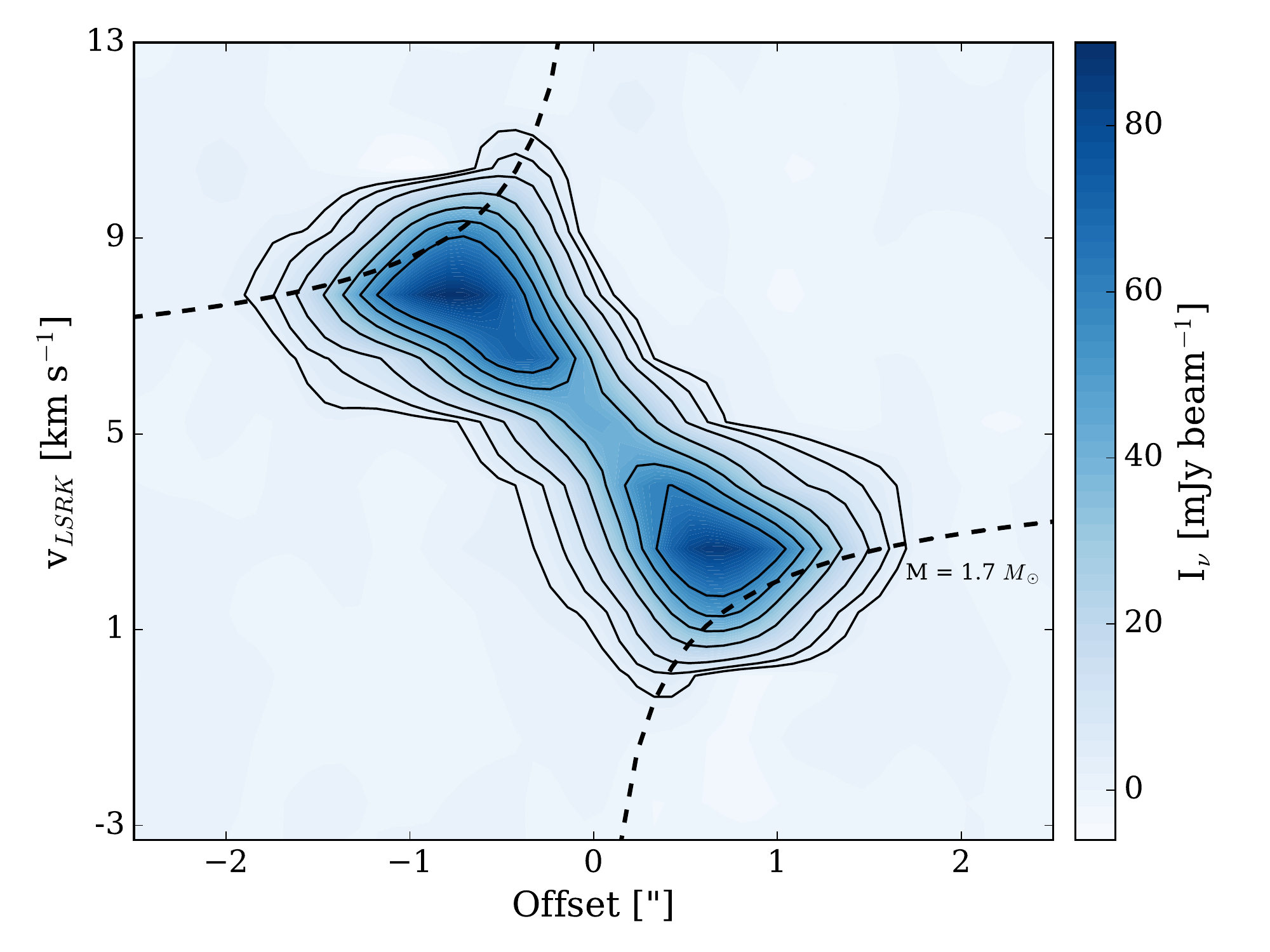}
  \end{center}
 \end{minipage}
\caption{\small The dust and $^{12}$CO J$=2-1$ gas emission originate from the same location in the HD~32297 disk at $\sim80$~AU, but the gas exhibits slower velocities than expected for Keplerian rotation.  \emph{(left)} The 1.3~mm continuum emission (color) with the 0th moment map overlaid as black contours in steps of $[3, 6, 12, 24, 36, 48, 64]\times1.0$~mJy~km~s$^{-1}$~beam$^{-1}$, the rms noise in the image, overlaid on the 1st moment.  
\emph{(right)}  The position-velocity diagram of the $^{12}$CO J$=2-1$ emission from the HD~32297 debris disk.  Contours are in steps of $[3, 6, 12, 24, 36, 48, 64]\times1.0$~mJy~km~s$^{-1}$~beam$^{-1}$, the rms noise in the image. The expected Keplerian velocity profile for a disk inclination of $83\fdg6$ (the best-fit to the dust continuum emission) and a stellar mass of 1.7~$M_\odot$ (typical for an A5V--A6V star) is shown by the dashed black line. 
}
\label{fig:pv}
\end{figure}

The HD~32297 system is one of three CO-rich debris disks around $30-40$~Myr-old A-type stars, a group which also includes HD~21997 (A3V, 69.6~pc) and 49 Ceti (A1V, 57.1~pc).  Our measured ALMA flux density makes HD~32297 one of the most CO-rich debris disks observed, comparable in integrated line flux to HD~21997 \citep{kos13} and $1.5\times$ that of 49 Ceti \citep{hug17}, after accounting for their different distances \cite[see][]{moor17}.  All three of these systems have a dust belt detected in addition to the observed gas emission.  Although, for HD~21997, the gas disk appears to have an inner radius closer to the central star than the dust.  In HD~32297, the dust and gas seem to be co-located, implying that both are likely created through collisions between the same planetesimals.  To illustrate this, Figure~\ref{fig:pv} (left) shows the ALMA $^{12}$CO J$=2-1$ emission overlaid as contours on the 1.3~mm continuum emission.  Unlike the other two systems, HD~32297 is the only one to display millimeter gas and dust emission from an extended halo component.  HD~32297 also has the latest spectral type of the group at A5V--A6V \cite[often misquoted in the literature as A0V, see][for further discussion]{deb09} with a mass of $\sim1.7$~$M_\odot$ and luminosity of $\sim8$~$L_\odot$, more comparable to the somewhat younger and more gas-poor $\beta$ Pictoris system (A6V, 19.8~pc, 23~Myr).

Figure~\ref{fig:pv} (right) shows the position-velocity diagram for the CO emission observed by ALMA.  For reference, the expected Keplerian rotation curve is overlaid as a black dashed line, assuming a stellar mass of $1.7$~$M_\odot$, typical for an A5V--A6V star, and an inclination of $83\fdg6$, the best-fit from our continuum modeling.  By eye, the rotation velocity of the gas at the peak locations appears to be $\pm2.5$~km~s$^{-1}$ from the systemic velocity, below the $\pm4.4$~km~s$^{-1}$ that is expected from Keplerian rotation for a ring at $\sim80$ AU (using a higher inclination of $\sim88\degr$, as suggested by scattered light observations, makes the fit even worse).  Although suggestive, this result is marginal given the low spectral resolution of our data.  With a channel width of only 1.3~km~s$^{-1}$, the observed velocity difference is only $\sim2\sigma$.  Intriguingly, sub-Keplerian rotation is also seen in the $\beta$ Pictoris debris disk.  By modeling CI and CII emission, \citep{cat18} derive a dynamical mass of 0.8~$M_\odot$, significantly below the assumed stellar mass of 1.75~$M_\odot$ \citep{crifo97} but comparable to the dynamical mass derived from Na I emission \citep{olof01}.

One explanation for sub-Keplerian rotation in debris disks could be radiation pressure, which acts opposite to a star's gravity, making the star appear less massive.  Mid to late A-type stars like HD~32297, however, do not emit significant UV radiation at the absorption wavelengths of CO, so this effect is likely have a negligible direct effect on neutral molecules. In the $\beta$ Pictoris disk, CO is quickly converted to C and O through photodissociation on timescales of $\sim50$~years.  \cite{fer06} suggest that these ionized species, which are directly affected by radiation pressure, are coupled into a single fluid through Coulomb interactions.  As a result, neutral molecules like CO could also become coupled and experience indirect braking due to radiation pressure. Given their similar spectral types, it is possible a similar explanation could be invoked in the case of HD~32297.  An alternative explanation was originally presented in \cite{wei77}, where a pressure gradient in a gas disk causes the gas to rotate more slowly than the free orbital velocity.  The deviation of the gas velocity from the Keplerian orbital velocity, $v_k$, is given by $\Delta v = v_k - v_g \approx -(\Delta g/2g)v_k$, where $g=v_k^2/r$ is the central gravity and $\Delta g=-nRT/\mu r$ is the residual gravity in a reference frame rotating with the gas.  Here, $n$ is the power law index of pressure, $P=P_0(r/r_0)^{-n}$, which we take to be 3, $R$ is the gas constant, $T$ is the gas temperature, and $\mu$ is the mean molecular weight.  Given the expected stellar mass ($1.7$~$M_\odot$), gas temperature (30~K), and radius of the HD~32297 debris disk, we estimate the magnitude of this effect to be $\sim0.1-0.5$~km~s$^{-1}$, not large enough to account for the $\sim2$~km~s$^{-1}$ observed difference in orbital velocity.  We note, however, that modifying the pressure gradient and mean molecular weight to account for a primordial or secondary origin of the gas could change this calculation significantly.  Although we do not discuss them at length here, two other possible explanations for the observed sub-Keplerian rotation of the gas could include non-uniform self-absorption or a stellar dynamical mass that disagrees with pre-main sequence models.

Circumstellar gas has been detected previously from the HD~32297 system.  \cite{red07} observed absorption from the Na I D doublet (5895.9242 and 5889.9510 \AA) with a column density of logN$_\text{NaI}\sim11.4$, the strongest Na~I absorption measured toward any nearby debris disk.  A $3.7\sigma$ detection of [CII] emission at 158~$\mu$m was also made with the Herschel PACS Spectrometer \citep{don13}.  Most recently, \cite{gre16} detected $^{12}$CO J$=2-1$ and tentatively J=$3-2$ emission with JCMT SCUBA-2 (taken several years earlier in 2013).  Their measured integrated intensity for $^{12}$CO J$=2-1$ is $3.7\pm0.4\times10^{-21}$~W~m$^{-2}$, giving a total CO mass of $1.3\times10^{-3}~M_\oplus$.  This result is $5\times$ ($7\sigma$) higher than our measured ALMA integrated intensity for the same transition of $7.91\pm0.08\times10^{-21}$~W~m$^{-2}$ with corresponding CO mass of $4.32\times10^{-4}~M_\oplus$.  The difference between these two measurements could easily stem from the lower signal-to-noise and resolution of the earlier JCMT observations.  However, \cite{gre16} also found a higher velocity for the gas, which implies a $2\times$ difference in the inner radius of the gas.  Given this, the CO production rate and location in the HD~32297 disk could be variable on the $\sim3$~year timescale between these two observations, with a higher production rate in 2013, although this would be challenging to explain given that the time between observations is much shorter than the orbital timescale for the system.

Given the relative wealth of gas detections from the HD~32297 debris disk, it is clear that this system offers us a unique opportunity to explore the properties of gas in main sequence circumstellar disks.  By combining our determinations of the total gas mass ($4.3\times10^{-3}~M_\oplus$ for CO and $4.3~M_\oplus$ for H$_2$, calculated by assuming an ISM-like CO/H$_2$ ratio of $10^{-4}$ expected only for a primordial system, see Section~\ref{sec:co}) and the total dust mass ($0.72\pm0.13~M_\oplus$, Section~\ref{sec:halo}), we can calculate the gas-to-dust ratio for the HD~32297 debris disk to be $6\times10^{-4}$ if CO dominates the gas mass (possibly higher if the gas is non-LTE, but still $<1$) or $6$ if H$_2$ is present in the disk.  As expected for an evolved system, the result assuming a primordial disk is significantly below the canonical value for the ISM of $\sim100$ \citep{boh78}, but is comparable to recent values determined from surveys of protoplanetary disks in much younger star forming regions \cite[e.g.,][]{ans16,mio17}.  Both results are also comparable to 49 Ceti, which has a gas to dust-ratio of either $0.008$ or $6$, depending on whether CO or H$_2$ makes the largest contribution to the gas mass, respectively \citep{hug17}.  \cite{kos13} calculate a much higher gas-to-dust ratio for HD~21997 of $300-700$.  This difference points to a hybrid scenario for HD~21997, where some primordial gas remains in the system, but dust is generated through secondary collisions between planetesimals.  For 49 Ceti and HD~32297, where the dust and gas are co-located, it seems more likely that both are created through secondary collisions between icy planetesimals.  Indeed, \cite{kral17} show that the mass of CO gas detected around HD~32297 is completely consistent with a secondary origin when considering non-LTE.  If this is the case, it may be illustrative to compare the CO-to-dust ratio in HD~32297 with Solar System comets; recent results from the Philae lander suggest that Comet 67P/Churyumov-Gerasimenko is more dust than ice with a volumetric ice-to-dust ratio of $0.4-2.6$ \citep{kof15}.  Additional observations with higher spectral resolution will be critical to confirm the origin of the gas in the HD~32297 system, as well as to examine the somewhat confusing gas dynamics and whether or not the CO production rate is indeed variable.

\section{Conclusions}
\label{sec:conc}

We present new ALMA observations of the HD~32297 and HD~61005 debris disks at 1.3~mm (230~GHz).  These images give us the highest angular resolution view to date of these two edge-on systems at millimeter wavelengths.  To analyze these data, we adopt an MCMC approach and fit directly to the millimeter visibilities.  The main results of this analysis are as follows.

\begin{enumerate}

    \item The 1.3~mm continuum emission from HD~32297 and HD~61005 is best described by a two-component power law model that includes both a planetesimal belt and an additional outer halo.  Simpler models with either a single power law or Gaussian leave significant residuals in the outer regions of the disk.  For HD~32297, the inner edge of the planetesimal belt is located at $78.5\pm8.1$~AU with an outer edge at $122\pm3$~AU, and a halo out to $440\pm32$~AU.  For HD~61005, the planetesimal belt extends between $41.9\pm0.9$~AU and $67.0\pm0.5$~AU, with a halo out to $188\pm8$~AU.  Although the best-fit models provide reasonably good fits to the data for both disks, we note some significant residuals at the $4-8\sigma$ level that could indicate additional asymmetric substructure not resolved by our current observations and likely make our quoted uncertainties underestimates.
    
    \item Our ALMA observations detect millimeter emission from an extended halo component in a debris disk.  Previously, the halos of the HD~32297 and HD~61005 debris disks were only seen in scattered light observations, implying that they were populated by small, micron-sized grains easily removed by interactions with the ISM or an interior planet. If, instead, millimeter grains are also present in these halos, we must reevaluate how such structures are created.  Since interactions with the ISM are only expected to affect small grains, these new observations favor sculpting via planetary perturbations.  It is also likely that the presence of gas in a disk, as is the case for HD~32297, will have an effect on the system's dynamics, since outflowing gas can entrain grains.
    
    \item  For both HD~32297 and HD~61005, the planetesimal belt component has a rising surface density gradient with power law indices of $x_1=2.23\pm0.61$ and $2.87\pm0.19$, respectively.  Similar to other systems that exhibit rising surface density gradients (e.g., AU Mic and $\epsilon$ Eridani), both belts are broad with fractional widths, $\Delta R/R> 0.4$.  These results contribute to a growing population of broad debris disks that exhibit rising surface density gradients, consistent with predictions of `self-stirred' models for debris disks where ongoing formation of small bodies at the inner edge of the disk triggers a collisional cascade. 
    
    \item We detect $^{12}$CO J$=2-1$ emission from the HD~32297 debris disk co-located with the dust continuum emission.  The integrated intensity is $7.91\pm0.08\times10^{-21}$~W~m$^{-2}$.  Assuming that the gas is optically thin and in LTE, this flux implies a total CO mass of $4.32\times10^{-4}~M_\oplus$, comparable to or more massive than other CO-rich debris disks.  Intriguingly, the measured ALMA integrated intensity is a factor of $5\times$ lower than previous measurements for the same system \citep{gre16}.  Although this difference could stem from the lower signal-to-noise and resolution of the earlier observations, it could also suggest variability in the CO production rate in this system.  In addition, the CO gas appears to exhibit sub-Keplerian rotation, although the significance of this result is unclear given the low spectral resolution of our data (channel width of 1.3~km~s$^{-1}$).
 
\end{enumerate}

The ALMA observations presented here provide new insights into the structure and dynamics of the HD~32297 and HD~61005 debris disks.  Ultimately, millimeter observations with higher angular and spectral resolution are required to confirm the dominant mechanism responsible for sculpting these systems.  A detection of additional substructure in either disk could reveal the presence and influence of a giant interior or exterior planet.

\vspace{1cm}
The authors thank David Rodriguez for leading the Cycle 1 ALMA proposal to observe HD~32297, Ian Czekala for helpful discussions regarding the HD~32297 gas detection, Gaspard Duch\^{e}ne for helpful discussions regarding the inclination of HD~32297, and the anonymous referee for comments that improved the clarity of the manuscript. M.A.M. acknowledges support from a National Science Foundation Astronomy and Astrophysics Postdoctoral Fellowship under Award No. AST-1701406. A.M.H. acknowledges support from NSF grant AST-1412647. This paper makes use of the following ALMA data: ADS/JAO.ALMA \#2015.1.00633.S and \#2012.1.00437.S. ALMA is a partnership of ESO (representing its member states), NSF (USA) and NINS (Japan), together with NRC (Canada) and NSC and ASIAA (Taiwan) and KASI (Republic of Korea), in cooperation with the Republic of Chile. The Joint ALMA Observatory is operated by ESO, AUI/NRAO and NAOJ. The National Radio Astronomy Observatory is a facility of the National Science Foundation operated under cooperative agreement by Associated Universities, Inc.  This work has also made use of data from the European Space Agency (ESA) mission {\it Gaia} (\url{https://www.cosmos.esa.int/gaia}), processed by the {\it Gaia} Data Processing and Analysis Consortium (DPAC, \url{https://www.cosmos.esa.int/web/gaia/dpac/consortium}). Funding for the DPAC has been provided by national institutions, in particular the institutions participating in the {\it Gaia} Multilateral Agreement.  


\bibliography{References.bib}

\pagebreak

\begin{deluxetable}{l l c c c c}
\tablecolumns{6}
\tabcolsep0.06in\footnotesize
\tabletypesize{\small}
\tablewidth{0pt}
\tablecaption{ALMA Observations \label{tab:obs}}
\tablehead{
\colhead{Target} &
\colhead{Observation} & 
\colhead{\# of } & 
\colhead{Projected} & 
\colhead{Precipitable Water} &
\colhead{Time on} \\
\colhead{ } &
\colhead{Date} & 
\colhead{Antennas} & 
\colhead{Baselines (m)} &
\colhead{Vapor (mm)} &
\colhead{Target (min)}
}
\startdata
HD 32297 & 2016 Jan 1 & 36 & $15-310$ & 1.26 & 22.2 \\
{} & 2016 Jun 21 & 36 & $15-704$ & 1.15 & 44.4 \\
\hline
HD 61005 & 2013 Dec 4$^\text{a}$ & 27 & $12-330$ & 2.99 & 49.0 \\
{} & 2014 Mar 20$^\text{a}$ & 31 & $12-335$ & 0.79 & 26.6 \\
{} & 2016 Jun 18 & 36 & $15-650$ & 0.73 & 44.4 
\enddata
\tablecomments{$^\text{a}$ Archival observations from ALMA project 2012.1.00437.S (PI: David Rodriguez)}
\end{deluxetable}

\begin{deluxetable}{l l c c}
\tablecolumns{3}
\tabcolsep0.1in\footnotesize
\tabletypesize{\small}
\tablewidth{0pt}
\tablecaption{Best-fit Model Parameters \label{tab:results}}
\tablehead{
\colhead{Parameter} & 
\colhead{Description} & 
\colhead{HD 32297} &
\colhead{HD 61005} \\
\colhead{} &
\colhead{} & 
\colhead{Best-fit} &
\colhead{Best-fit}
}
\startdata
$F_\text{belt}$ & Planetesimal belt flux density [mJy] & $3.04\pm0.21$ & $4.82\pm0.29$ \\
$F_\text{halo}$ & Halo flux density [mJy] & $0.63\pm0.26$ & $1.54\pm0.48$ \\
$R_\text{in}$ & Planetesimal belt inner edge [AU] & $78.5\pm8.1$ & $41.9\pm0.9$  \\
$R_\text{halo}$ & Planetesimal belt outer edge and disk halo inner edge [AU] & $122\pm3$ & $67.0\pm0.5$ \\
$R_\text{out}$ & Halo outer edge [AU] & $440\pm32$ & $188\pm8$ \\
$x_1$ & Planetesimal belt power law gradient & $2.23\pm0.61$ & $2.87\pm0.19$ \\
$x_2$ & Halo power law gradient & $-6.20\pm0.30$ & $-5.52\pm0.14$ \\
$i$ & Disk inclination [$\degr$] & $83.6 (+4.6, -0.4)^\text{a}$ & $85.6\pm0.1$ \\
$PA$ & Disk position angle [$\degr$] & $47.9\pm0.2$ & $70.3\pm0.1$
\enddata
\tablecomments{$^\text{a}$ We estimate an additional $5\%$ uncertainty towards higher inclinations from a simple $\chi^2$ analysis in the image plane (see Section~\ref{sec:compare_hd32297} for discussion)}
\end{deluxetable}

\end{document}